\documentclass[12pt,fleqn]{article}
\usepackage{graphicx}

\usepackage{latexsym}

\usepackage{amsmath}
\usepackage{amsthm}
\usepackage{amssymb}
\usepackage{amsfonts}

\numberwithin{equation}{section}

\newcommand{\be}{\begin{equation}} 
\newcommand{\ee}{\end{equation} \par \noindent}

\newcommand{\rf}[1]{(\ref{#1})}

\newcommand{\ods}{\par \vspace{1.5ex} \par}

\newcommand{\ba}{\begin{array}}
\newcommand{\bac}{\begin{array}{c}}
\newcommand{\bal}{\begin{array}{l}}
\newcommand{\ea}{\end{array}}

\newcommand{\ex}{\\[1ex]}   
\newcommand{\exx}{\\[2ex]}

\newcommand{\mc}{\multicolumn}

\newcommand{\Frac}[2]{ \frac{\displaystyle #1}{\displaystyle #2} }

\newcommand{\const}{{\rm const}}
\newcommand{\diag}{{\rm diag}}
\newcommand{\Tr}{{\rm Tr}}

\newcommand{\N}{{\cal N}}
\newcommand{\R}{{\mathbb R}}
\newcommand{\C}{{\mathbb C}}

\newcommand{\la}{\lambda}

\newcommand{\ph}{\varphi}

\newcommand{\m}{\left( \ba{c}}
\newcommand{\mm}{\left( \ba{cc}}

\newcommand{\miv}{\left( \ba{rrrr}}
\newcommand{\ema}{\ea \right)}

\newcommand{\ket}[1]{\mbox{$\mid \! #1 \ \rangle$}}      
\newcommand{\bra}[1]{\mbox{$\langle \ #1 \! \mid$}}
\newcommand{\scal}[2]{\mbox{$\langle #1 \! \mid #2 \rangle $}} 

\newcommand{\tm}{\! \times \!} 

\newcommand{\dis}{\displaystyle }
\newcommand{\no}{\noindent}

\newtheorem{prop}{Proposition}[section]
\newtheorem{Th}[prop]{Theorem}   
\newtheorem{lem}[prop]{Lemma}
\newtheorem{rem}[prop]{Remark}
\newtheorem{cor}[prop]{Corollary}

\begin{document}

\title{{\bf Algebraic construction of   
the Darboux matrix revisited} }
\author{Jan L. Cie\'sli\'nski\thanks{\footnotesize e-mail: \tt janek\,@\,alpha.uwb.edu.pl } 
\\ {\footnotesize Uniwersytet w Bia{\l}ymstoku, 
Wydzia{\l} Fizyki, 
   15-424 Bia\l ystok, ul.\ Lipowa 41, Poland}
}

\date{}

\maketitle 

\begin{abstract} 
We present algebraic construction of Darboux matrices for 1+1-dimensional    
integrable systems of nonlinear partial differential equations with a special stress on the nonisospectral  case. 
We discuss 
different approaches to the Darboux-B\"acklund transformation, based on different $\lambda$-dependencies of the Darboux matrix: polynomial, sum of partial fractions, 
or the transfer matrix form. 
We derive symmetric $N$-soliton formulas in the general case. 
The matrix spectral parameter and dressing actions in loop groups are also discussed. 
We describe reductions to twisted loop groups, unitary reductions, 
the matrix Lax pair for the KdV equation and reductions   
of chiral models (harmonic maps) to $SU(n)$ and to Grassmann spaces.  
We show that in the KdV case the nilpotent Darboux matrix generates the 
binary Darboux transformation. 
The paper is intended as a review of known results (usually presented in a novel context) 
but some new results are included as well, e.g., general compact formulas for $N$-soliton surfaces and linear and bilinear constraints on the nonisospectral Lax pair matrices which are preserved by Darboux transformations. 
\end{abstract}

\no {\it PACS Numbers}: 02.30.Ik, 03.50.Kk, 05.45Yv. 

\no {\it MSC 2000}: 37K35, 37K30, 37K25, 35Q53, 22E67.

\no {\it Keywords}: integrable systems, Darboux-B\"acklund transformation, Darboux matrix, dressing method, 
loop groups, reduction group, nonisospectral linear problems, invariants of Darboux transformations, $N$-soliton surfaces, chiral models, KdV equation

\

\section{Introduction}

A 1+1-dimensional integrable system can be considered as integrability
conditions for a {\it linear problem} (a system of linear partial
differential equations defined by two matrices containing the {\it spectral
parameter}), see for instance \cite{ZMNP}.  The Darboux-B\"acklund  transform is a gauge-like transformation (defined by the {\it Darboux matrix}) which preserves  the form of the linear problem \cite{Ci-dbt,CLMRW,GHZ,LRS,SZ}. All approaches to the construction of Darboux matrices originate in the dressing method \cite{ZMNP,Szab,ZS1,ZS2}. 

The paper is intended as a presentation of Darboux-B\"acklund transformations 
from a unified perspective, first presented in \cite{Ci-ChSF,Ci-dbt}.  
The construction of
the Darboux matrix is divided into two stages. First, we uniquely characterize the considered linear problem in terms of algebraic constraints (the divisor of poles, 
loop group reductions and other algebraic properties, e.g., linear and biblinear constraints). Then, 
we construct the Darboux matrix preserving all these constraints. Using general theorems, including those from the present paper, one may  
construct the Darboux matrix in a way which is 
almost algorithmic.

The paper is intended as a review of known results but some new results are also included. 
We discuss in detail elementary Darboux transformation (Darboux matrix which has a single simple zero), symmetric formulas for Darboux matrices and soliton surfaces (in the general case), and loop group reductions for polynomial Darboux matrices. Two examples are discussed in detail: the 
Korteweg-de Vries equation and chiral models (harmonic maps). 

The part which seems to be most original contains the description of linear and 
bilinear invariants of Darboux transformations. 
We prove that  multilinear constraints introduced in \cite{Ci-dbt}
are invariant  with respect to the polynomial 
Darboux transformation (also in the nonisospectral case).  Taking them into account 
we can avoid some cumbersome calculations, our construction assumes a more elegant form and, 
last but not least, we do not need any assumptions concerning boundary conditions. 
 
Another important aim of this paper is to show similarities and even an equivalence between different  algebraic approaches to the construction of the Darboux matrix. This is a novelty in itself because sometimes it is difficult to notice connections between different methods. The existing monographs, even the recent ones, focus on a chosen single approach, compare \cite{DL,GHZ,MS,MNS,ZMNP,RS-dbt}. 
\ods\ods

We consider a nonlinear system of partial differential equations
which is equivalent to the compatibility conditions
\be \label{cc}
    U_\mu,_\nu - U_\nu,_\mu + [U_\mu, U_\nu] = 0 \ , 
\quad (1 \leqslant \mu < \nu \leqslant m) \ ,
\ee
for the following system of linear equations (known as the Lax pair, at least in the case of two independent variables)
\be \label{ZS}
   \Psi,_\nu = U_\nu \Psi \ , \quad (\nu=1,\ldots,m) \ ,
\ee
where  $n \times n$ matrices $U_\nu$ depend on $x^1,\ldots,x^m$ and on the so called spectral parameter $\lambda$ (and, as usual, $\Psi,_\nu = \partial \Psi/\partial x^\nu$, etc.). We assume that $\Psi$ is also a matrix (the fundamental solution of the linear system \rf{ZS}). 
We fix our attention on the case $m=2$ (although most results hold for any $m$) 
and shortly denote by $x$ the set of all variables, i.e., $x = (x^1, \ldots, x^m)$.

\subsection{Isospectral and nonisospectral Lax pairs} 

Let us recall that the most important characteristic of the matrices
$U_1$, $U_2$ is their dependence on the spectral parameter $\lambda$.
In the typical case $U_\nu$ are rational with respect to $\lambda$. Actually we 
will consider a more general situation. We assume that the Lax pair 
is rational with respect to $\lambda$, and 
\begin{itemize}
\item ``isospectral case'':  $\lambda$ is a constant parameter,
\item ``non-isospectral'' case: 
\be  \label{Lnu}
 \lambda,_\nu = L_\nu (x,\lambda) \ , \quad (\nu=1,\ldots,m) \ ,
\ee
where $L_\nu$ are given functions, rational with respect to $\lambda$ (this case reduces to the isospectral one for $L_\nu (x,\lambda) \equiv 0$). 
\end{itemize}

\begin{rem}  \label{Lambda}
The differential equations \rf{Lnu} are of the first order, so their solution  
$\lambda = \Lambda (x, \zeta)$ depends on a constant of integration $\zeta$ which plays the role of the constant spectral parameter.
\end{rem}

 The solution of the system \rf{Lnu} exists provided that compatibility conditions hold,  
for more details see \cite{Ci-dbt}.  
In general $\Lambda = \Lambda (x,\zeta)$ is an implicit function, although in many special cases explicit expression for $\Lambda$ can be found, compare \cite{BZM,Ci-dbt,Sh}).

\subsection{The Darboux-B\"acklund transformation}

The application of the dressing method   to  generate  new solutions of
nonlinear equations ``coded'' in \rf{cc} consists in the following (see
\cite{ZMNP,ZM-ZETF,ZS2}).  Suppose that we are able to construct a gauge-like 
transformation $\tilde{\Psi} = D \Psi$ (where $D=D(x,\lambda)$ will be called the
Darboux matrix) such that the structure of  matrices $\tilde{U}_\nu$, 
\be    \label{DBT}
     \tilde{U}_\nu = D ,_\nu D ^{-1} + D U_\nu D ^{-1} \ ,  \qquad  (\nu=1,\ldots,m) \ ,
\ee
  is  identical  with the structure of the matrices $U_\nu$. The soliton fields  entering  $U_\nu$  are 
replaced by some new fields which, obviously, have to satisfy  the 
nonlinear system \rf{cc} as well. 

\begin{rem}  \label{divis}
The Darboux transformation should preserve divisors of poles (i.e., poles and their multiplicities) of matrices $U_\nu$. This is the most important structural property of \  $U_\nu$ to be preserved. The second important property is the so called reduction group, see Section~\ref{redu}. 
\end{rem}

      For any pair  of solutions of \rf{cc} one can ``compute''
$D:=\tilde{\Psi}\Psi^{-1}$. The crucial point is, however, to express $D$ solely
by the wave function $\Psi$ because only then one can use $D$ to construct new
solutions. Such $D$ is known  as the Darboux matrix \cite{LRS,Mat,MS}.
	The Darboux matrix defines an explicit map $S \mapsto S$,
	where $S$ is the set of solutions of the linear problem \rf{ZS}.
The construction of the Darboux matrix 
is based on the important observation:

\begin{rem}
The Darboux matrix can be expressed in an algebraic way by the original
 wave function $\Psi$. 
\end{rem} 

By the ``original wave function'' we mean one before the transformation.  In
fact, it is rather difficult to find special solutions of the linear problem.
Usually very limited number of cases is available.  However, knowing any 
solution $\Psi = \Psi (x, \lambda)$ and the Darboux matrix one can generate 
a sequence of explicit solutions. Starting from the trivial background ($x$-independent and mutually commuting $U_\nu$) we usually get the so called soliton solutions.

\subsection{Equivalent Darboux matrices}

It is quite natural to consider as equivalent Darboux matrices
which produce exactly the same transformation \rf{DBT} of matrices 
$U_\nu$ of a given linear problem.  

\begin{rem}  \label{rem-C}
The linear problem \rf{ZS} is invariant under transformations
$\Psi \mapsto \Psi C_0$ (for any constant nondegenerate matrix $C_0 = C_0 (\lambda)$).
\end{rem}

Therefore Darboux matrices $D$ and $D'$ are equivalent if there exists a
matrix $C$ such that $D\Psi = D'\Psi C$  (for any $\Psi$). Thus $C$ 
should commute with $\Psi$ what, in practice, means that $C = f (\lambda) \in \C$. 

\begin{rem}  \label{rem-equiv}
   The matrix $D'=f(\lambda)D$, where $f$ is a complex function of $\lambda$ only, 
	is equivalent to $D$.
\end{rem}

\subsection{Soliton surfaces approach}
\label{sec-Sym}

Given a solution $\Psi = \Psi (x, \lambda)$, where $\lambda$ depends on $x$ and $\zeta$, we define a new object $F$ 
by the so called Sym-Tafel (or Sym) formula: 
\be  \label{Sym} 
    F = \Psi^{-1} \Psi,_\zeta  \ , 
\ee
If  $\Psi$  assumes values in a matrix Lie group $G$, than (for any 
fixed $\zeta$) $F$ describes an immersion (a ``soliton surface'') into the corresponding Lie algebra 
\cite{Sy1,Sym}. Soliton surfaces are a natural frame to unify a variety of 
different physical models like soliton fields, strings, vortices, chiral models and spin models \cite{Sy2}. In the framework of the soliton surfaces approach one can reconstruct many integrable cases known from the classical differential geometry \cite{Bob,Ci-izot,Ci-PWN,Sym}. 
The Darboux-B\"acklund transformation for soliton surfaces reads
\be  \label{DBT-F}
   \tilde F =  F + \Psi^{-1} D^{-1} D,_\zeta \Psi 
\ee
where $D,_\zeta = \lambda,_\zeta D,_\lambda$.  The equivalent Darboux matrices yield the same soliton surfaces. Indeed, if we take $D' = f D$, then
\be  \label{F+f}
  \tilde F = F + \Psi^{-1} D'^{-1} D',_{\zeta} \Psi = 
F + \frac{f,_{\zeta}}{f} + \Psi^{-1} D^{-1} D,_\zeta \Psi \ , 
\ee
i.e., surfaces corresponding to $D$ and $D'$ differ by the constant  
$(\ln f),_\zeta$. 

\ods
In order to illustrate usefulness of the geometric approach we present the  
following theorem \cite{Ci-hyper}.

\begin{Th}
We assume that $U_1, U_2$ are linear combinations of $1$, $\lambda$ and $\lambda^{-1}$,   
with $x$-dependent $su(2)$-valued coefficients, and $U_\nu (-\lambda)= E_0 U_\nu (\lambda) E_0^{-1}$ (where $E_0 \in su(2)$ is a constant matrix).  
Then $F$ given by the Sym formula \rf{Sym} is (in the isospectral case) 
a pseudospherical (i.e., of negative Gaussian curvature) surface immersed in $su(2) \simeq \R^3$. In the nonisospectral case the same assumptions yield the so called Bianchi surfaces.  
\end{Th}

We point out that surprisingly few assumptions (restrictions) on the spectral problem leads to the very important class of pseudospherical surfaces. 
It is easy to assure the preservation of these restrictions by the Darboux transformation. 

Darboux transformations usually preserve many other constraints (e.g., linear and bilinear invariants discussed in Section~\ref{invariants}) what  leads to the preservation of some geometric characteristics (e.g, curvature lines) and to a specific choice of coordinates and other auxiliary parameters.

\section{Binary Darboux matrix}

In this paper by the binary Darboux matrix we mean one pole matrix 
with non-degenerate normalization 
\be  \label{Dbin}
 D = \N \left( I + \frac{\lambda_1 - \mu_1}{\lambda- \lambda_1} P \right) \ , 
\quad  P^2 = P \ , \quad  \det \N \neq 0 \ , 
\ee
such that its inverse has the same form:
\be  \label{Dbin-inv}
 D^{-1} = \left( I + \frac{\mu_1 - \lambda_1}{\lambda- \mu_1} P \right) \N^{-1} \ . 
\ee
Here $\lambda_1, \mu_1$ are complex parameters (which can depend on $x$ in the nonisospectral case), $P=P(x)$ is a projector matrix ($P^2 = P$), and $\N = \N(x)$ is the so called normalization matrix.  

\subsection{Binary or elementary?}

The name ``binary'' for Darboux matrices of the form \rf{Dbin} is rather tentative, 
because binary Darboux transformations were introduced in another context 
(compare \cite{MS,ZMS}). The ``classical'' binary transformation corresponds to 
the degenerate case of \rf{Dbin} when $\mu_1 \rightarrow \lambda_1$ (see Section~\ref{binary}), i.e.,  
\be  \label{M-nil}
D = \N \left( I + \frac{M}{\lambda-\lambda_1} \right) \ , \quad 
D^{-1} = \left( I - \frac{M}{\lambda-\lambda_1} \right) \N^{-1} \ ,
\ee
where $M^2 = 0$ (the so called nilpotent case, see \cite{Ci-dbt}). 
Therefore, we use this notion in an extended sense. However, it seems to be compatible with understanding binary Darboux transformation as a composition of an elementary Darboux transformation  
and a Darboux transformation of the adjoint linear problem \cite{MS,OS,SR}. 
In the case of the Zakharov-Shabat spectral problems \rf{ZS} the adjoint 
spectral problem is given by  
\be  \label{ZS-ad}
 - \Phi,_\nu = \Phi U_\nu \ , 
\ee
and one can easily check that $\Phi = \Psi^{-1}$ solves the adjoint spectral problem. The general solution of \rf{ZS-ad} is $\Phi = \Psi^{-1} C$, where $C$ is a constant (i.e., $x$-independent) matrix. We will see in Section~\ref{sec-ZM} that the binary Darboux matrix can be expressed by a pair of solutions: one solves the spectral problem \rf{ZS}, and 
the second one solves the adjoint problem \rf{ZS-ad}.

The matrix \rf{Dbin} is equivalent to the linear in $\lambda$ matrix $\hat D$
\be  \label{lin-bin}
\hat D = \N \left( \lambda - \lambda_1  + (\lambda_1 - \mu_1) P \right) \ . 
\ee
Darboux matrices linear in $\lambda$  are sometimes referred to as ``elementary'', see  \cite{RS-dbt}. Indeed, iterating such transformations we can get any Darboux transformation with nondegenerate normalization. 
However, we reserve the name ``elementary''  for 
matrices which are not only linear in $\lambda$ but have a single zero (see Section~\ref{elem}), or even a single simple zero. The polynomial form  \rf{lin-bin} of the binary Darboux matrix has two 
zeros:  $\lambda_1, \mu_1$.  The sum of their multiplicities is $n$. Therefore these zeros are simple only in the case $n=2$. 

\subsection{Sufficient conditions for the projector}

Assuming that $U_\nu$ are regular (holomorphic) at $\lambda=\lambda_1$ and $\lambda=\mu_1$,
and demanding that $\tilde U_\nu$ (expressed by \rf{DBT}) 
have no poles at $\lambda=\lambda_1$ and $\lambda=\mu_1$ as well, we get the following conditions (for vanishing the corresponding residues), compare \cite{Ci-dbt}:   
\be \bal     \label{12}
     P \circ \left( - \partial_\nu   + U_\nu(\lambda_1) \right) \circ (I-P) = 0  \ ,  \\[2ex]
     (I-P)\circ \left(  - \partial_\nu   + U_\nu(\mu_1) \right)  \circ P = 0 \ ,
\ea \ee
\be \bal \label{12-lamy}
\lambda_1,_\nu = L_\nu (x,\lambda_1) \ , \qquad \mu_1,_\nu = L_\nu (x,\mu_1) \ ,
\ea \ee   
where the circles mean composition of linear operators and $L_\nu$ are defined 
by \rf{Lnu}. Note that for any operators $A, B$ we have:
\be
A\circ B=0 \quad \Rightarrow \quad {\rm im}  B \subset \ker A \ .
\ee
Indeed, $(A\circ B)\ph=0$ for any vector $\ph$, i.e. $A(B\ph)=0$ what  means exactly that $B\ph\in \ker A$. On the other hand, any element of ${\rm im}  B$ is of the form $B\ph$.

\begin{rem}
The assumption that  $U_\nu$ are regular at $\lambda=\lambda_1$ and $\lambda=\mu_1$ (assumed throughout this paper) is essential. Relaxing this requirement we can get  solutions 
different from those obtained by the standard Darboux-B\"ack\-lund transformation. 
Solutions of this kind ({\it unitons}) have been found in the case of harmonic maps into Lie groups \cite{Uh}, see also \cite{GHZ}.
\end{rem}

If the system \rf{Lnu} has the general solution $\lambda = \Lambda (x,\zeta)$, then the equations \rf{12-lamy} can be solved in terms of the function $\Lambda$:
\be  \label{zety}
  \lambda_1 = \Lambda (x,\zeta_1) \ , \qquad \mu_1 = \Lambda (x, \zeta'_1) \ ,
\ee 
where $\zeta_1, \zeta'_1$ are constant parameters, compare \cite{BZ,NK}. 
Taking into account $\ker P = {\rm im} (I - P)$, we easily show that the system  \rf{12} is equivalent to: 
\be  \bal \label{subset}
  \left( - \partial_\nu   + U_\nu (\lambda_1) \right) \ker P  \subset \ker P \ , \exx
   \left( - \partial_\nu   + U_\nu (\mu_1) \right) {\rm im}  P  \subset {\rm im}  P \ .
\ea \ee              

Now we easily see that the conditions \rf{subset} are satisfied by
the projector defined by the Zakharov-Shabat formulas 
(compare \cite{ZMNP,ZM-ZETF,ZS2}):
\be  \label{14}
     \ker  P = \Psi (\lambda_1) V_{ker} \ , \quad
     {\rm im}   P  = \Psi (\mu_1) V_{im} \ ,
\ee
where $\Psi (\lambda_1)=\Psi (x, \lambda_1)$, $\Psi (\mu_1)=\Psi
(x, \mu_1)$ and  $V_{ker}$ and $V_{im}$ are
constant vector spaces such that $V_{ker} \oplus V_{im} = V$.
 Indeed,  in this case, by virtue of \rf{ZS}, the left-hand sides of \rf{subset} are simply 
equal to zero. 

Taking into account that any projector $P$ 
can be expressed explicitly by its kernel and image,  
$P = ({\rm im}  P, 0) ({\rm im}  P, \ker P)^{-1}$, 
we can summarize the above discussion as follows.

\begin{prop}  \label{ZS prop}
The transformation \rf{DBT} with $D$ given by \rf{Dbin},
where  
\be  \label{zsp}
P=(\Psi(\mu_1) V_{im},\ 0)(\Psi(\mu_1) V_{im},\ \Psi(\lambda_1)
V_{ker})^{-1} \ , 
\ee
preserves the divisors of poles of matrices $U_\nu$.
\end{prop}

The formula \rf{zsp} yields a sufficient condition for $P$ to 
generate the Darboux matrix. It is interesting to find also necessary conditions.
Therefore, we will try to obtain the most general  solution  of \rf{subset}. 

\subsection{The general form of the binary Darboux matrix}
\label{general}

It is convenient to represent vector spaces in a matrix form. Namely,
if $w_1,\ldots,w_k$ span a vector space $V$, then we can identify $V$ with 
the matrix
\be  \label{kolumny}
    V = (w_1, \ldots, w_k) \ . 
\ee
This matrix has $k$ columns ($w_1,\ldots,w_k$) and $n$ rows ($n=\dim V$).

Note that because of the freedom in choosing a basis in the vector space
there are many matrices representing the same vector space. 
If $a_{ij}$ are coefficients of a $k\tm k$ non-degenerate matrix $A$,  
then the vectors
\[
   w'_j = \sum_{i=1}^{k} w_i a_{ij}  \ ,
\]
form another basis in $V$ which can be represented by the matrix
\[
   V' = (w'_1, \ldots, w'_k) = V A  \ .
\] 
The matrices $V$ and $V'$ (for any non-degenerate $A$) represent the
same vector space and, in this context, are considered as equivalent ones.

The space of $k$-dimensional subspaces of an $n$-dimensional vector space over
$\C$ is known as Grassmannian $G_{k,n}(\C)$ (of course, considering real vector
spaces we have real Grassmannian $G_{k,n}(\R)$).  The elements of the
Grassmannian are classes of equivalence of $k\tm n$ matrices with respect to the
equivalence relation: $V \simeq V'$ if there exists $k\tm k$ matrix $A$ 
($\det A \neq 0$) such that $V'=VA$.

Therefore using the same notation for the vector space and the matrix 
representing it, one should remember about this equivalence. In particular,
in order to show that some vector spaces $W$ and $V$ are identical, one has to consider
the equation $W=VA$ with an arbitrary non-degenerate $A$. In the similar way
one can check whether $W$ is a subspace of $V$ (of course a necessary 
requirement is $\dim V \geq \dim W$). 

\ods
\begin{prop}  \label{W subset V}
Let $V, W$ be vector spaces, $k' = \dim W \leq \dim V = k$. Then $W \subset V$
if and only if there exists $k\tm k$ matrix $B$ such that $W = V B$.
\end{prop}

\no Proof: If $W \subset V$, then there exists a
basis of V such that its first $k'$ vectors span $W$. We represent $V$ by 
vectors of this basis, i.e., we choose $A$ such that $w'_1,\ldots,w'_{k'}$ span
$W$. Finally, we put $B = A\
\diag(1,\ldots,1,0,\ldots,0)$. \hfill $\Box$
\ods

Note that formally $W$ and $V$ belong (in general)  to different Grassmannians.
But if $\det B = 0$, then the columns of $VB$ are linearly dependent and 
$VB$ can be treated as an element of a Grassmannian of lower dimension.
\ods 

We proceed to solving the system \rf{subset}. 
Assuming  $\det \Psi(\lambda_1) \neq 0$ and $\det \Psi(\mu_1) \neq 0$, we can always put $\ker P = \Psi (\lambda_1) V$,  ${\rm im}  P = \Psi (\mu_1) W$,
where $W$, $V$ are some vector spaces (in general $x$-dependent). 
Substituting to \rf{subset} we have:
$ \Psi(\lambda_1) V,_\nu \subset \Psi(\lambda_1) V$, 
$\Psi(\mu_1) W,_\nu \subset \Psi(\mu_1) W$. 
Hence, 
\[    V,_\nu \subset V \ , \qquad W,_\nu \subset W \ . \]
By Proposition~\ref{W subset V}, we rewrite $V,_\nu \subset V$ as
$V,_\nu = VB_\nu$ for some $B_\nu$ (which have to satisfy appropriate compatibility conditions), 
and analogical equations for $W$. 
Taking into account the freedom of changing the basis
when changing $x$: $V'=VA$  ($\det A \neq 0$),  we obtain: 
\[  (V'A^{-1}),_\nu = V' A^{-1} B_\nu \]
Therefore, choosing $A$ such that 
\be  \label{A-1}
(A^{-1}),_\nu = A^{-1} B_\nu \ , 
\ee
we obtain $V',_\nu=0$, i.e. there exists an $x$-independent basis in $V$ (the same conclusion holds for $W$). The solution of \rf{A-1} exists because $B_\nu$ satisfy the compatibility conditions  mentioned above.  Thus we have shown that the formulas \rf{14} give the most general solution of
\rf{subset}.

\section{Polynomial Darboux matrices: general case}

In this paper we consider only rational Darboux matrices   
($n\times n$ matrices with coefficients which are rational functions of $\lambda$). 

\begin{rem}
In the isospectral case, every rational Darboux matrix is equivalent to a polynomial Darboux matrix
\be  \label{Dpol} 
 \hat D = \sum_{k=0}^N  T_{k} (x)  \lambda^{N-k} \  . 
\ee
Indeed, it is enough to multiply given $D$ by the least common multiple of all 
denominators. The obtained polynomial will be denoted by $\hat D$. 
\end{rem}

Another equivalent form of $D$ is a polynomial in $\lambda^{-1}$, obtained from $\hat D (\lambda)$ by dividing it by $\lambda^N$.  In some cases this polynomial  is more convenient that $\hat D$ because it is analytic at $\lambda = \infty$.  

In the nonisospectral case the least common multiple of all denominators depends on $x$. Therefore, any rational Darboux matrix is equivalent to some polynomial matrix up to a scalar $x$-dependent factor.

\subsection{The determinant of the Darboux matrix}

The trace of a quadratic matrix is defined as the sum of
diagonal elements of this matrix. 
Both the trace and the determinant are invariant with respect to  
similarity transformations: $\Tr(BAB^{-1}) = \Tr A$, \ $\det (BAB^{-1}) = \det A$.

\begin{Th}[Liouville]  \label{Liouville}
If $\Psi,_\nu = U_\nu \Psi$, where $\nu$ is fixed and $U_\nu = U_\nu (x)$ is given, then 
\be  \label{Liou}
(\det\Psi),_\nu = \Tr\,U_\nu \det\Psi \ . 
\ee
\end{Th}
\no This theorem is well known as the Liouville theorem on Wronskians, 
see, for instance, \cite{Arn}.

Applying the Liouville theorem to the Darboux transform  $\tilde \Psi \equiv D \Psi$ we get \ $(\det D \det \Psi ),_\nu = \Tr \tilde U_\nu \det D \det \Psi$. Hence, 
using once more \rf{Liou}, we obtain:
\be   \label{DUU} 
    \frac{(\det D),_\nu}{\det D} =  \Tr(\tilde{U}_\nu) - \Tr(U_\nu) \ .
\ee

\begin{rem} \label{rem-detconst}
We usually consider traceless linear problems ($\Tr\
U_\nu =  0$ for $\nu = 1,\ldots,m$). In such case 
   $\det D$ has to be  constant (i.e., $\det D$ does not depend on \ $x$). 
Therefore, in the isospectral (and traceless) case $\det D$ can depend only 
on $\lambda$ and all its zeros are constants. 
\end{rem}

In the nonisospectral case the situation is more complicated because $\lambda$ depends on $x$. However, it is still possible to obtain a strong general result characterizing zeros of $\det D$. 

\begin{Th}  \label{Th-nonis}
 We consider a polynomial Darboux matrix \ $\hat D$ for a nonisospectral 
linear problem \rf{ZS} with $\lambda$ satisfying \rf{Lnu}.  
If \ $\det \hat D(\lambda_k) = 0$ and matrices $U_\nu$ are regular at $\lambda_k$, then  
\be  \label{lambdak}
\lambda_k,_\nu = L_\nu (x,\lambda_k) \ , 
\ee
i.e., $\lambda_k = \Lambda (x, \zeta_k)$, where $\zeta_k = \const$.  
\end{Th}

\no Proof: The determinant of the polynomial $\hat D (\lambda)$ has a finite number of roots ($x$-dependent, in general). We denote them by $\lambda_k$, $k=1,\ldots,K$, and their multiplicities by $m_k$. Note that $m_1 + m_2+\ldots m_K = n N$, where $N$ is the degree of the polynomial $\hat D (\lambda)$ and $n$ is the order of the matrix $\hat D$. Thus
\be
 \det \hat D (\lambda) = h \prod_{k=1}^K (\lambda-\lambda_k)^{m_k} \ , 
\ee
where $h = h (x)$ and $\lambda_k = \lambda_k (x)$. Taking into account \rf{Lnu} we compute 
\be  \label{dD} 
  \frac{(\det D),_\nu}{\det D} = 
\frac{h,_\nu}{h} + \sum_{k=1}^{K} \left( m_k \frac{ L_\nu (x,\lambda) - \lambda_k,_\nu }{\lambda - \lambda_k} \right)  \ . 
\ee
The equation \rf{DUU} with $U_\nu$ regular at $\lambda_k$ implies that the right-hand side of \rf{dD} should have no poles. Therefore residua of \rf{dD} 
at $\lambda=\lambda_k$ vanish what implies \rf{lambdak}. The $x$-dependence of $\lambda_k$ follows from Remark~\ref{Lambda}. \hfill  $\Box$\ods

The regularity of $U_\nu$ at $\lambda=\lambda_k$ is assumed throughout this paper. 
If we allow that some $\lambda_k$ coincides with a singularity of $U_\nu$, then 
 the $x$-dependence of $\lambda_k$ in principle can be different from \rf{lambdak} and we get  an  additional freedom. 

\subsection{Neugebauer's approach} \label{Neugebauer}

A simple but quite general method to construct polynomial Darboux-B\"acklund
transformations has been proposed by Neugebauer and his collaborators \cite{MNS,NK,NM}, see also \cite{HS,RS-dbt}.
We are going to find conditions on polynomial ${\hat D}$
implying that divisors of poles of $\tilde{U}_\nu$ and $U_\nu$ coincide (compare 
Remark~\ref{divis}). From $\tilde \Psi,_\nu = \tilde U_\nu \tilde \Psi$ we get
\be  \label{DBT-T}
\tilde{U}_\nu = \frac{\tilde \Psi,_\nu (\lambda) \tilde \Psi^{c} (\lambda) } {\det \tilde\Psi (\lambda) } = \frac{1}{\det \hat D} ({\hat D},_\nu {\hat D}^{c} + {\hat D} U_\nu {\hat D}^{c}) \ ,
\ee
where by ${\hat D}^{c}$ we denote the matrix of cofactors of ${\hat D}$. 
Obviously ${\hat D}^{c}$ is also a polynomial in $\lambda$.

If $U_\nu$ are rational functions of $\lambda$, then $\tilde U_\nu$ given 
by \rf{DBT-T} are rational as well (because ${\hat D}$ and ${\hat D}^{-1}$ are rational).
Therefore the only candidates for poles of $\tilde{U}_\nu$ are  poles  
of $U_\nu$ and zeros of $\det {\hat D}$ (i.e., $\lambda_k$). 
The necessary condition for the regularity of $\tilde U_\nu$ at $\lambda = \lambda_k$ is
\be \label{res-zero}
\tilde\Psi,_\nu (\lambda_k) \tilde\Psi^{c}(\lambda_k) = 0 \ .
\ee
If $\lambda_k$ is a simple zero of $\det \hat D (\lambda)$, then the condition 
\rf{res-zero} is also sufficient. 
\ods
Following \cite{MNS}, we will find another, more constructive, characterization of the condition 
\rf{res-zero}. 
If $\det {\hat D}(\lambda_k) = 0$, then we have also
\be  \label{detpsi0} 
 \det \tilde{\Psi}(\lambda_k) = 0 \ 
\ee
(because $\tilde \Psi (\lambda) = \hat D (\lambda) \Psi (\lambda)$). 
We assume that the function $\Psi (\lambda)$ (known as a ``background solution'' or a ``seed solution'') is non-degenerate at $\lambda=\lambda_k$. 

As a consequence of \rf{detpsi0}, the equation 
$\tilde{\Psi}(\lambda_k) p_k = 0$ has a non-zero solution $p_k \in \C^n$ 
(where, in principle, $p_k$ can depend on $x$).  Then, we compute:
\[ 
\tilde \Psi,_\nu (\lambda_k) p_k = \tilde U_\nu (\lambda_k) \tilde \Psi (\lambda_k) p_k = 0 \ ,
\]
where we took into account that $\tilde \Psi (\lambda)$ satisfies \rf{ZS}.   
Thus we have:
\be
\tilde \Psi (\lambda_k) p_k  = 
\tilde \Psi,_\nu (\lambda_k) p_k  = 0 \ , 
\ee
what  implies \rf{res-zero}, as one can see from the following fact of linear algebra (\cite{MNS}, see also \cite{HS}).

\begin{lem}  \label{XY^{c}}
 Let us consider two degenerated matrices $X$ and $Y$. Suppose that there
exists a vector $p$ such that 
$ X p = 0$  and  $ Y p = 0$. Then: $YX^{c}=0$.
\end{lem}

\no Proof: 
Let us perform computations in a basis $(e_1, \ldots, e_n)$ such that
$e_1 \equiv p$. Then all elements of the first column of matrices $X$, $Y$ are equal to zero. Thus, using the definition of the cofactor, we easily see that 
the rows of $Y^{c}$ (except the first row) have all entries equal to zero. 
Hence, $XY^{c}$ obviously yields zero. 
\hfill  $\Box$

\begin{lem}  \label{const-eigen}
The vector $p_k$ such that \ $\tilde{\Psi}(\lambda_k) p_k = 0$ \ is defined up to 
a scalar factor. If $\lambda_k$ is a simple zero, then we can choose this multiplier in such a way that\ $p_k = \const$. 
\end{lem}

\no Proof: We differentiate the equation defining $p_k$:\  
$\tilde{\Psi},_\nu (\lambda_k) p_k + \tilde \Psi (\lambda_k) p_k,_\nu = 0$. Hence, $\tilde \Psi (\lambda_k) p_k,_\nu = 0$, which means that $p_k,_\nu$ is proportional to $p_k$ (provided that  $\lambda_k$ is a simple zero of $\det \hat D (\lambda)$). Thus $p_k,_\nu = f_{k\nu} p_k$, where $f_{k\nu}$ are some 
scalar functions. From the identity $p_k,_{\nu \mu} \equiv p_k,_{\mu\nu}$ 
it follows that $f_{k\nu},_\mu = f_{k\mu},_\nu$. Therefore, there exists $\ph_k$ such that $f_{k\nu} = \ph_k,_\nu$. Hence, $p_k e^{- \ph_k} $ does not depend on $x$. 
\hfill  $\Box$

\begin{cor}  \label{cor-Dar} 
Polynomial Darboux matrix \rf{Dpol} 
can be constructed as follows. In the isospectral case we choose $Nn$ pairwise different complex numbers $\lambda_1,\lambda_2,\ldots,\lambda_{Nn}$ and $N n$ constant $\C^n$-vectors $p_1, p_2,\ldots, p_{Nn}$. We also choose the matrix  $T_0$ (``normalization  matrix''), $\det T_0 \neq 0$. 
Matrix coefficients $T_1,\ldots, T_N$ are computed from
\be  \label{TFi}
     {\hat D}(\lambda_k) \Psi(\lambda_k) p_k = 0 \ , \quad (k=1,\ldots, nN) \ ,
\ee
where $\Psi (\lambda)$ is given (``seed solution''). In the nonisospectral case we choose constants $\zeta_1,\ldots,\zeta_{N n}$ and use \rf{lambdak}. 
\end{cor}

For a fixed $k$ the equation \rf{TFi} consists of $n$ scalar equations. 
Thus we have a system of $n^2 N$ equations for $N$ unknown matrices $n\times n$. 
In the generic case such system should have a unique solution. 

 The freedom in choosing $T_0$ corresponds  to  a gauge transformation. Note that an identical situation has place in the case of the binary  Darboux matrix, where $\N$ is, in general, undetermined. Usually it is sufficient to put $T_0 = I$ (``canonical normalization''). If this choice leads to a contradiction (i.e., the Darboux matrix with the canonical normalization does not exist), then we may relax  this assumption and search for Darboux matrices with more general normalization. 

The case $\det T_0 = 0$ can be treated in a similar way but with one exception: the total number of zeros is smaller than $N n$. As an example of such situation we will present elementary Darboux matrices, see Section~\ref{elem}.

\subsection{Explicit multisoliton formulas}
\label{ex-multi}

Let us introduce the notation
\be
   \ph_k := \Psi(\lambda_k) p_k \ ,
\ee 
where $\ph_k \in \R^n$ are column vectors. 
We assume $\det T_0 \neq 0$ and denote
\be
   \theta_j:=T_0^{-1} T_j \ , \qquad (j=1,\ldots,N) \ , 
\ee
where $T_j$ are defined by \rf{Dpol}. The equations \rf{TFi} read:
\be
     \left( \lambda_k^N + \sum_{j=1}^N \lambda_k^{N-j} \theta_j \right) \ph_k = 0 \ ,
     \qquad (k=1,\ldots,M) \ ,
\ee
where $M = n N$. After the transposition we get: 
\be
 \sum_{j=1}^N \ph_k^T \theta_j^T \lambda_k^{N-j} = - \ph_k^T \lambda_k^N \ .
\ee
It is convenient to solve these equations in the matrix form: 
\be  \label{Ck} 
\m \theta_1^T \\ \theta_2^T \\ \cdots \\ \theta_N^T \ema = - 
\miv \lambda_1^{N-1} \ph_1^T & \ldots & \lambda_1 \ph_1^T & \ph_1^T \\
     \lambda_2^{N-1} \ph_2^T & \ldots & \lambda_2 \ph_2^T & \ph_2^T \\
	  \mc{4}{c}{\dotfill} \\
     \lambda_{M}^{N-1} \ph_{M}^T & \ldots & \lambda_{M} \ph_{M}^T & \ph_{M}^T \ema^{-1}
	  \m \ph_1^T \lambda_1^N \\ \ph_2^T \lambda_2^N \\ \ldots \\ \ph_{M}^T \lambda_{M}^N \ema
      \ . 
\ee
Usually, in practical applications, one uses Cramer's rule to express $\theta_k$ in terms of determinants, compare \cite{NM,PS,SMN}. 

Having coefficients $T_k$ we can apply the Darboux transformation to 
Lax pairs of prescribed form. As an illustrative example we present the simplest but very important case (linear in $\lambda$):  
\be
  U_1 = u_0 \lambda + u_1 \ , 
\ee
The equation \rf{DBT} for $\nu=1$, i.e., $\tilde U_1 D = D U_1 + D,_1$, yields:
\be
\left( \tilde u_0 \lambda + \tilde u_1 \right) \sum_{k=0}^N \lambda^{N-k} T_k = \sum_{k=0}^N \lambda^{N-k} T_k \left( u_0 \lambda + u_1 \right) + \sum_{k=0}^N \lambda^{N-k} T_k,_1  \ .
\ee
Considering coefficients by $\lambda^{N+1}$ and $\lambda^N$, we get explicit formulas for the transformed fields $\tilde u_1$ and $\tilde u_0$: 
\be  \label{tilde-u01}
\tilde u_0 = T_0 u_0 T_0^{-1} \ , \quad 
\tilde u_1 = T_0 u_1  T_0^{-1} + [ T_1 T_0^{-1}, \tilde u_0] + T_0,_1 T_0^{-1} \ . 
\ee
In the classical AKNS case $u_0 = i\sigma_3 \equiv  \diag (i,-i)$ and it is sufficient to take the canonical normalization $T_0 = I$. Therefore, we get 
\be
\tilde u_0 = u_0 \ , \quad  \tilde u_1 = u_1 + [ T_1, i\sigma_3 ] \ , 
\ee
where $T_1 = \theta_1$ can be explicitly computed from \rf{Ck}, compare \cite{NM}.

\section{Elementary Darboux matrix}
\label{elem}

The elementary Darboux matrix is linear in $\lambda$ and its determinant has just a single simple zero. This case is mentioned by Its \cite{Its} and discussed in more detail in, for instance,  \cite{DL,LU}.
An obvious way to produce matrices of this type is 
to take matrices with a single entry linear in $\lambda$ and all other entries $\lambda$-independent.  In this paper we confine ourselves to elementary Darboux matrices for $n=2$. They can be represented in the form
\be
 D = \N \mm \lambda - \lambda_1 & 0 \\ - \alpha & 1 \ema {\cal M}
\ee
where $\N, {\cal M}$ do not depend on $\lambda$. As a simple exercise (compare Corollary~\ref{cor-Dar}) we can express the coefficient $\alpha$ by $\Psi$ evaluated at $\lambda_1$, namely:
\be
 \alpha = \frac{\eta_1}{\xi_1} \ , \quad  \m \xi_1 \\ \eta_1 \ema = 
{\cal M} \Psi (\lambda_1) p_1 \ ,
\ee
 where $p_1$ is a constant vector.

\subsection{Binary Darboux matrix as a superposition of elementary transformations} 

\begin{Th}  \label{Th-super}
In the case $n=2$ any binary Darboux transformation is a superposition of two elementary Darboux transformations. 
\end{Th}

\no Proof: We will show that 
\be  \label{superel}
  D = \N_2 \mm 1 & - \beta \\ 0 & \lambda - \lambda_2 \ema  \N_1^{-1} \N_1 
\mm \lambda - \lambda_1 & 0 \\ -\alpha & 1 \ema {\cal M} \ ,
\ee
 is a binary Darboux matrix ($\N_1, \N_2, {\cal M}$ are non-degenerate matrices which do not depend on $\lambda$). First, performing the multiplication 
in \rf{superel}, we get  
\be
 D = \N \left( \lambda - \lambda_1 + 
 (\lambda_1 - \lambda_2)  P  
 \right) 
\ee
where
\be \ba{l} 
\N = \N_2 \mm 1 & 0 \\ -\alpha & 1 \ema {\cal M}  \ , \\[3ex] \dis 
P = \frac{1}{\Delta \lambda} {\cal M}^{-1} \mm \alpha \beta &  -\beta \\  \alpha (\alpha \beta - \Delta \lambda) &  \Delta \lambda - \alpha \beta \ema {\cal M} \ , 
\ea \ee
and  $\Delta \lambda = \lambda_1 - \lambda_2$. Then, we easily check that 
$P^2 = P$.

\ods

The coefficients $\alpha, \beta$ can be expressed by $\Psi$ evaluated at $\lambda_1, \lambda_2$. Indeed, denoting
\be
 \Psi (\lambda_k) p_k  = \mm \xi_k \\ \eta_k \ema \ , 
\ee
and using equations \rf{TFi}, we obtain
\be  \label{albet} 
   \alpha = \frac{\eta_1}{\xi_1} \ , \quad 
\beta = \frac{\xi_1 \xi_2 \Delta \lambda}{\xi_2 \eta_1 - \eta_2 \xi_1} \ . 
\ee
The projector $P$ reads
\be  \label{Pkset}
  P = \frac{1}{\xi_1 \eta_2 - \eta_1 \xi_2} {\cal M}^{-1} \mm - \eta_1 \xi_2 & \xi_1 \xi_2 \\ 
- \eta_1 \eta_2 & \eta_2 \xi_1 \ema {\cal M} \ .
\ee
If ${\cal M} = I$, then  $\Psi (\lambda_1) p_1 \in \ker P$ and $\Psi (\lambda_2) p_2 \in {\rm im} P$. Therefore, the binary Darboux matrix with $P$ given by 
\rf{Pkset} is a superposition of elementary transformations \rf{superel} with 
${\cal M} = I$ and $\alpha, \beta$ given by \rf{albet}.  
\hfill $\Box$\ods

\subsection{KdV equation}

The Darboux transformation for the famous Korteweg-de Vries equation is 
almost always presented in the scalar case, see \cite{MS}.
The matrix approach is less convenient. However, having in mind 
a pedagogical motivation, we are going to show in detail that 
the matrix construction works also in that case. It is interesting, that in 
this paper we do not need the ``KdV reality condition'' (usually used in earlier papers, compare \cite{Ci-dbt,GHZ,TU}). 

The standard scalar Lax pair for KdV equation consists of the Sturm-Liouville-Schr\"odinger spectral problem and the second equation defining the time evolution of the wave function:
\be \label{KdV-scal-Lax}
- \psi,_{11} + u  \psi =  \lambda \psi \ , \qquad 
\psi,_2 = - 4\psi,_{111} + 6 u \psi,_1 + 3 u,_1 \psi \ .
\ee
The compatibility conditions $\psi,_{112} = \psi,_{211}$ yield the KdV equation
\be  \label{KdV}
u,_2 - 6 u u,_1 + u,_{111} = 0 \ .
\ee
The Lax pair \rf{KdV-scal-Lax} can be transformed, in a standard way, to the matrix form
\be  \ba{l}  \label{Lax-KdV} 
 \Psi,_1 = \left( \ba{cc} 0 & 1 \\ u - \lambda & 0 \ea \right) \Psi 
 \ , \\[2ex]
\Psi,_2 = \left( \ba{cc} - u,_1 & 2 u + 4 \lambda \\ -4\lambda^2 + 2 u\lambda + 2 u^2 
- u,_{11} & u,_1 \ea \right) \Psi \ , 
\ea \ee
where
\be  \label{Psi-psi}
\Psi = \left( {\vec \psi} , {\vec \phi} \right) \ , \quad 
\vec \psi = \left( \ba{l} \psi \\ \psi,_1 \ea \right)  \ , \quad 
\vec \phi = \left( \ba{l} \phi \\ \phi,_1 \ea \right) \ ,
\ee
and $\psi$, $\phi$ are linearly independent solutions 
 of \rf{KdV-scal-Lax}. 

\begin{lem}  \label{lem-abc}
Suppose that 
\be \label{UV-KdV}
U = \mm 0 & 1 \\ u-\lambda & 0 \ema \ , \quad V = \mm 0 & 4\lambda \\ 
2 \lambda u - 4 \lambda^2 & 0 \ema + \mm - a & b \\ c & a \ema \ ,
\ee
where \ $u, a, b, c$ \ do not depend on $\lambda$. Then, the compatibility conditions $U,_2 - V,_1 + [U, V] = 0$ uniquely yield:
\be  \label{abc}
a = u,_1 \ , \quad b = 2 u \ , \quad c = 2 u^2 - u,_{11} \ . 
\ee
i.e., $U, V$ given by \rf{UV-KdV} are identical with the Lax pair \rf{Lax-KdV} for the KdV equation. 
\end{lem}

\no Proof is straightforward: compatibility conditions reduce to \rf{KdV} and \rf{abc}.

\subsection{Elementary Darboux matrix and the classical Darboux transformation}
\label{clas-Dar}

We will compute the action of the elementary Darboux transformation in the KdV case, compare \cite{GHZ}. We assume
\be  \label{Del}
 D = \N \left( \ba{cc} \lambda - \lambda_1 & 0 \\ - \alpha & 1 \ea \right) 
\ee
where $\N$ ($\det \N \neq 0$) does not depend on $\lambda$ and $\alpha$ is a function to be expressed by $\Psi (\lambda_1)$, namely
\be  \label{con-1}
 D (\lambda_1) \Psi (\lambda_1) p_1 = 0 \ , 
\ee
where $p_1$ is a constant vector. We denote 
\be
     \m \xi_1 \\ \eta_1 \ema = \Psi (\lambda_1) p_1 
\ee
The constraint \rf{con-1} (with $D$ given by \rf{Del}) is 
equivalent to 
\be  \label{alpha}
   \alpha = \frac{\eta_1}{\xi_1} = \frac{\hat \psi_1,_1}{\hat \psi_1} \ ,  
\ee
where $\hat \psi_1$ satisfies \rf{KdV-scal-Lax} with $\lambda=\lambda_1$ (i.e., $\hat \psi_1$ is a linear combination of $\psi_1$ and $\phi_1$). 
The function $\alpha$ satisfies the following system of Riccati equations:
\be \ba{l}  \label{Ric}
\alpha,_1 = u - \lambda_1 - \alpha^2 \ , \\[2ex] 
\alpha,_2 = (2 u^2 - u,_{11} + 2 u \lambda_1 - 4 \lambda_1^2) + 2 u,_1 \alpha - ( 2 u + 4 \lambda_1)   \alpha^2 \ ,  
\ea \ee
which can be obtained directly from \rf{Lax-KdV}. 

The elementary Darboux transformation for $U, V$ (i.e., the formulas \rf{DBT} with $D$ given by \rf{Del}) reads:
\be \ba{l} \label{KdV-trans1} \dis
\tilde U = \frac{M_1}{\lambda - \lambda_1} 
+ \lambda \N \mm 0 & 1 \\ 0 & 0 \ema \N^{-1} - \N \mm - \alpha & \lambda_1 \\ 1 &  \alpha \ema \N^{-1} + \N,_1 \N^{-1} 
\ , \\[3ex]  \dis
\tilde V = \frac{M_2}{\lambda - \lambda_1}  
+ \left( 4 \lambda^2 + b \lambda \right) \N \mm 0 & 1 \\ 0 & 0 \ema \N^{-1} 
- 4 \lambda  \N \mm - \alpha &  \lambda_1 \\ 1 &  \alpha \ema \N^{-1} 
+ \tilde V_0  , 
\ea \ee
where 
\be \ba{l} \dis
M_1 = \left( u - \lambda_1 - \alpha^2 - \alpha,_1 \right)
\N \mm 0 & 0 \\ 1 & 0 \ema \N^{-1} 
\ , \\[3ex] \dis
M_2 = \left( -\alpha,_2 + c + 2 a \alpha - b \alpha^2 - 4 \lambda_1^2 + 2\lambda_1 (u - 2 \alpha^2) \right)  \N \mm 0 & 0 \\ 1 & 0 \ema \N^{-1} \ ,
\ea \ee
and $\tilde V_0$ does not depend on $\lambda$ (its explicit form follows from Lemma~\ref{lem-abc} and, therefore, is automatically preserved by the Darboux transformation). 

The necessary condition for the Darboux transformation is vanishing of residua  
$M_1, M_2$ (what  is equivalent to \rf{con-1} and, as a consequence, to the Riccati equations \rf{Ric}). 

In order to assure the Darboux invariance of the coefficients by $\lambda$ in $U$ and by $\lambda^2$  in $V$ we have to impose some constraints on the normalization matrix $\N$ (compare \cite{LSW-nor}), namely
\be
 \N \mm 0 & 1 \\ 0 & 0 \ema \N^{-1} = \mm 0 & 0 \\ -1 & 0 \ema \ ,
\ee
what  implies the following form of $\N$:
\be
 \N = f \mm 0 & 1 \\ -1 & - \gamma \ema \ , 
\ee
where $f, \gamma$ are functions of $x$. 
Now, the transformation \rf{KdV-trans1} becomes
\be  \label{KdV-trans2} \ba{l} \dis
\tilde U =  \mm \gamma-\alpha & 1 \\ \tilde u - \lambda & \alpha-\gamma \ema + \frac{f,_1}{f} \mm 1 & 0 \\ 0 & 1 \ema \ , \\[3ex] \dis 
\tilde V = 4 \lambda \mm \gamma-\alpha & 1 \\ \tilde v - \lambda & \alpha-\gamma \ema + \frac{f,_2}{f} \mm 1 & 0 \\ 0 & 1 \ema + \tilde V_0 \ , 
\ea \ee
where
\be \ba{l}  \label{tildes}
\tilde u = \lambda_1 + 2\gamma\alpha-\gamma^2 - \gamma,_1 \ , \\[2ex]
\tilde v = \lambda_1 + 2\gamma\alpha-\gamma^2 - \frac{1}{4} b \ .  
\ea \ee
Comparing \rf{KdV-trans2} with \rf{UV-KdV} we find the remaining constraints on the form of the Darboux matrix:
\be
 f =\const \ , \quad \gamma = \alpha \ , \quad 2 \tilde v = \tilde u \ .
\ee
We assume $f = -1$. By virtue of \rf{abc} $b = 2 u$, and we easily verify that the  constraint $2 \tilde v = \tilde u$ coincides with the first Riccati equation \rf{Ric}. 

\begin{cor} 
The elementary Darboux matrix for the KdV equation is given by
\be  \label{D-elem}
 D = \mm 0 & 1 \\ -1 & -\alpha \ema \mm \lambda- \lambda_1 & 0 \\ - \alpha & 1 \ema = \mm - \alpha & 1 \\ \alpha^2 - \lambda + \lambda_1 & -\alpha \ema \ , 
\ee
where $\alpha$ is computed from \rf{con-1}, see also \rf{alpha}. 
\end{cor}

The transformation of $u$ can be obtained from \rf{Ric} and \rf{tildes}, 
see \rf{class-Dar}. 
Taking into account \rf{Psi-psi} we get the transformation for $\psi$. 

\begin{cor}
The elementary Darboux matrix \rf{D-elem} generates the classical Darboux transformation:
\be \ba{l} \label{class-Dar}
\tilde \psi = \psi,_1 - \alpha \psi \equiv \psi,_1 - (\ln \hat \psi_1),_1 \psi \ , \\[2ex] 
\tilde u = u - 2 \alpha,_1 \equiv u - 2 (\ln \hat \psi_1 ),_{11} \ ,
\ea \ee
where $\hat \psi_1 = \hat \psi (x, \lambda_1)$ satisfies \rf{KdV-scal-Lax}. 
\end{cor}

\no Formulas \rf{class-Dar} were first obtained by Gaston Darboux \cite{Dar}, see also \cite{MS}. 

\begin{prop} 
$D \equiv D_{\alpha, \lambda_1}$ given by \rf{D-elem} has the following 
properties: $D_{\alpha, \lambda_1}^{-1}$ is equivalent to 
$D_{-\alpha, \lambda_1}$ and \ 
$D_{\beta, \lambda_2} D_{\alpha, \lambda_1} = \N ( \lambda - \lambda_1 + M)$, where $M = (\lambda_1-\lambda_2) P$ (and $P^2=P$) for $\lambda_2 \neq \lambda_1$ and $M^2 = 0$ for $\lambda_2 = \lambda_1$.
\end{prop}

\no Proof: by straightforward computation. First, 
$D_{-\alpha, \lambda_1} = (\lambda_1-\lambda) D_{\alpha, \lambda_1}^{-1}$. 
Then, 
\be
D_{\beta, \lambda_2} D_{\alpha, \lambda_1} = \mm -1 & 0 \\ \alpha + \beta & -1 \ema 
\left( \lambda - \lambda_1 + M \right) 
\ee
where 
\be
M = \mm - \alpha (\alpha + \beta)   & \alpha + \beta  \\ 
\alpha (\lambda_1 - \lambda_2) -  \alpha^2 (\alpha + \beta)  &  
 \alpha (\alpha + \beta)  - (\lambda_1 - \lambda_2) \ema ,
\ee 
and we easily verify that $M^2 = (\lambda_1 - \lambda_2) M$, which means that (for $\lambda_2 \neq \lambda_1$)  
$M = (\lambda_1 - \lambda_2) P$ (where $P^2 = P$), compare \rf{lin-bin}. 
\hfill $\Box$\ods

\subsection{Nilpotent Darboux matrix and classical binary Darboux transformation}
\label{binary}

Let us consider the Darboux matrix of the form \rf{M-nil}. In the case $n=2$ the nilpotent matrix $M$ ($M^2=0$) can be parameterized as
\be
 M = g \mm - \sigma & 1 \\ - \sigma^2 & \sigma \ema 
\ee
where $g, \sigma$ are some functions. 

Considering the transformation \rf{DBT} we have to demand that  
$\tilde U_\nu$ are regular at $\lambda_1$ by cancelling the pole of second order at $\lambda=\lambda_1$. We get two conditions:
\be \ba{l}  \label{MM} 
M,_\nu + [M, U_\nu (\lambda_1)] - M U'_\nu (\lambda_1) M = 0 \ , \\[2ex]
M,_\nu M + M U_\nu  (\lambda_1) M = 0 \ , 
\ea \ee
where the prime denotes differentiation with respect to $\lambda$. The second set of equations turns out to be a consequence of the first equations (it is enough to multiply them by $M$ from the right). 

The classical binary Darboux transformation is usually defined only for the 
time-independent spectral problem \cite{MS}. Therefore, in order to show that the considered transformation \rf{M-nil} coincides with the classical binary transformation it is sufficient to confine ourselves to $\nu=1$. The equations \rf{MM} (for $\nu=1$) can be rewriten in terms of $g, \sigma$: 
\be \ba{l}  \label{threee}
g,_1 - 2 \sigma g + g^2 = 0 \ , \\[2ex]
g,_1 \sigma + g \sigma,_1 + \sigma g^2 - g (u - \lambda_1 + \sigma^2) = 0 \ , \\[2ex] 
g,_1 \sigma^2 + 2 g \sigma \sigma,_1 + \sigma^2 g^2 - 2 g \sigma (u - \lambda_1) = 0 \ , 
\ea \ee 
Using the first equation we can reduce the last two equations to:
\be
  \sigma,_1 + \sigma^2 - u + \lambda_1 = 0 \ .
\ee
Therefore 
\be  \label{sigma}
  \sigma = \frac{\hat \psi_1,_1}{\hat \psi_1} \ ,
\ee
where $\hat \psi_1$ satisfies the first equation of \rf{KdV-scal-Lax} for $\lambda=\lambda_1$,  
compare \rf{alpha}, \rf{Ric}. 
Taking into account \rf{MM} we rewrite  \rf{DBT} for $U_1 \equiv U = u_0 \lambda + u_1$  as
\be  \label{dbtM} 
\tilde u_0 = \N u_0 \N^{-1} \ , \quad 
\tilde u_1 = \N,_1 \N^{-1} + \N \left( u_1 + [ M, u_0 ] \right) \N^{-1} \ .
\ee
In the KdV case, see \rf{UV-KdV}, 
the first equation of \rf{dbtM} is satisfied for  
\be
\N = \mm 1 & 0 \\ \gamma & 1 \ema \ . 
\ee
Then, the second equation of \rf{dbtM} reduces to:
\be
\tilde u = u + \gamma,_1 + 2 \sigma \gamma +\gamma^2 \ , \qquad g = - \gamma \ .
\ee
Taking into account the first equation of \rf{threee} we finally get
get 
\be  \label{uga}
\tilde u = u + 2 \gamma,_1 \ , \quad  \gamma,_1 - 2 \sigma \gamma - \gamma^2 = 0 \ ,
\ee
where $\sigma$ is given by \rf{sigma}.  Therefore, the last equation is equivalent to:
\be
  \frac{\partial}{\partial x} \left( \frac{\hat \psi_1^2}{\gamma} \right) = - \hat \psi_1^2 \ ,
\ee
which means that 
\be
\gamma = \frac{\hat \psi_1^2}{c_0 - \int \hat \psi_1^2 } \ , \qquad 
\tilde u = u - \frac{\partial^2}{\partial x^2} \ln \left|  c_0 - \int \hat \psi_1^2 \right| \ ,
\ee
where $c_0$ is a constant of integration. The last formula coincides with the 
classical binary Darboux transformation for the Sturm-Liouville-Schr\"odinger 
spectral problem \cite{MS}. 

\begin{cor}
The nilpotent Darboux matrix \rf{M-nil} generates the classical binary Darboux transformation. 
\end{cor} 

The ``second'' binary Darboux transformation, introduced in \cite{ZMS}, corresponds to the choice $c_0=1$.

\section{Fractional form of the Darboux matrix}

Another popular representation of the Darboux matrix (with nondegenerate normalization) 
is decomposition into partial fractions  
\cite{ZMNP,ZM-ZETF,ZM-CMP,ZS2}:
\be \ba{l} \dis  \label{D-frac}
   D = \N(I + \Frac{A_1}{\lambda-\lambda_1} + \ldots + \frac{A_N}{\lambda-\lambda_N}) \ , \\[3ex] \dis
	  D^{-1} = (I + \Frac{B_1}{\lambda-\mu_1} + \ldots + \frac{B_N}{\lambda-\mu_N}) \N^{-1} \ .
\ea \ee 
In principle the numbers of poles of $D$ and $D^{-1}$ could be different but 
here, following other papers, we assume the ``symmetric'' case \rf{D-frac}.  

We will denote by $D_0$ the Darboux matrix in the fractional form with the canonical normalization (in other words, $D = \N D_0$). The form \rf{D-frac} of $D$ and $D^{-1}$ imposes restrictions 
on $A_k$ and $B_k$ implied by equations $D D^{-1}=I$ and $D^{-1} D = I$, 
see \rf{DD-1}.

 Multiplying $D$ by the least common multiple of the denominators we obtain the equivalent polynomial form $\hat D (\lambda)$ (a polynomial of $N$th degree). The determinant  $\det \hat D (\lambda)$ is a polynomial of degree $N n$ vanishing at poles of $D$ and $D^{-1}$, i.e., at $\lambda=\lambda_k$ and $\lambda=\mu_k$ ($k=1,\ldots,N$). The sum of multiplicities of all zeros 
 of $\det \hat D (\lambda)$ equals $N n$. Therefore, for $n=2$ all zeros are simple, while for $n>2$ some of them have to be multiple zeros. 

The fractional form is convenient in the case of some reductions 
(e.g., orthogonal or unitary), when the  eigenvalues $\lambda_k$ $(k=1,\ldots,nN)$ can be naturally divided into pairs $\lambda_k, \mu_k$.

\subsection{Zakharov-Mikhailov's approach}
\label{sec-ZM}

We start from fractional representation of the Darboux matrix 
\rf{D-frac}, where 
$A_k, B_k$ have to satisfy constraints resulting from 
the condition $D D^{-1}=I$: 
\be  \ba{l} \label{DD-1}  \dis
 A_k \left( I + \sum_{j=1}^N \frac{B_j}{\lambda_k-\mu_j} \right) = 0 \ , \qquad
 \left( I + \sum_{j=1}^N \frac{A_j}{\mu_k-\lambda_j} \right) B_k = 0 \ ,
\\[4ex] \dis
\left( I + \sum_{j=1}^N \frac{B_j}{\lambda_k-\mu_j} \right) A_k = 0 \ , \qquad
 B_k \left( I + \sum_{j=1}^N \frac{A_j}{\mu_k-\lambda_j} \right) = 0 \ ,
\ea \ee  \par \no 
$(k=1,\ldots,N)$. 
We assume the nonisospectral case and 
demand that $\tilde U_\nu$ defined by \rf{DBT} have the same 
form as $U_\nu$. In particular, it means that the right-hand sides 
of \rf{DBT} have no poles. 
Equating to zero the residua at $\lambda=\lambda_j$ and at $\lambda=\mu_k$, we get
\be \ba{l} \dis  \label{DD-2}
\left( A_j,_\nu + A_j U (\lambda_j) \right) \left( I + \sum_{i=1}^N \frac{B_i}{\lambda_j - \mu_i} \right) + \left( L_\nu (\lambda_j) - \lambda_j,_\nu 
\right) \sum_{i=1}^N \frac{A_j B_i}{(\lambda_j - \mu_i)^2} = 0  , 
\\[4ex] \dis
\left( I + \sum_{i=1}^N  \frac{A_i}{\mu_k - \lambda_i}\right) \left(  U (\mu_k) B_k  - B_k,_\nu \right) - \left( L_\nu (\mu_k) - \mu_k,_\nu \right) 
\sum_{i=1}^N \frac{A_i B_k}{(\mu_k-\lambda_i)^2} = 0 ,
\ea \ee
for $j,k=1,\ldots,N$. Multiplying first equations by $A_j$ from the right and the second equations by $B_k$ from the left, and then using \rf{DD-1}, we obtain
\be 
\left( L_\nu (\lambda_j) - \lambda_j,_\nu 
\right) \sum_{i=1}^N \frac{A_j B_i}{(\lambda_j - \mu_i)^2} = 0  , 
\quad
\left( L_\nu (\mu_k) - \mu_k,_\nu \right) 
\sum_{i=1}^N \frac{A_i B_k}{(\mu_k-\lambda_i)^2} = 0 ,
\ee
which is satisfied when \rf{lambdak} (and similar equations for $\mu_k$) hold.  
Note that we derived here a proposition analogical to Theorem~\ref{Th-nonis}. 

In order to solve the system \rf{DD-1}, \rf{DD-2} we assume \rf{lambdak} and represent 
$A_k$, $B_k$ as follows:
\be  \label{Dir-AB}
 A_k = \ket{s_k} \bra{a_k} \ , \qquad  B_k = \ket{b_k} \bra{q_k}
\ee
where $\ket{s_k}$, $\ket{b_k}$ are matrices built of linearly independent 
$n$-component column vectors and $\bra{q_k}$,  $\bra{a_k}$  are matrices built of linearly independent $n$-component row vectors. In other words, all these  matrices have maximal rank. In particular, $\ket{s_k}$ and $\bra{a_k}$ have the same rank (denoted by ${\rm rk} A_k$) but (in general) different than the rank of $\ket{b_k}$ and  $\bra{q_k}$ (denoted by ${\rm rk} B_k$).  
Using the notation \rf{Dir-AB} we rewrite equations \rf{DD-1} and \rf{DD-2}  as follows:
\be \ba{l}  \dis  \label{DD-1-Dir}
\ket{s_k} \bra{a_k}  D_0^{-1} (\lambda_k) = 0 \ , \qquad 
D_0 (\mu_k) \ket{b_k} \bra{p_k} = 0 \ , \\[3ex]  \dis
D_0^{-1} (\lambda_k) \ket{s_k} \bra{a_k}  = 0 \ , \qquad 
\ket{b_k} \bra{p_k} D_0 (\mu_k) = 0 \ ,
\ea \ee
\be \ba{l} \dis  \label{DD-2-Dir}
\Big( \ket{s_k},_\nu \bra{a_k} + \ket{s_k} \bra{a_k},_\nu + 
\ket{s_k} \bra{a_k} U_\nu (\lambda_k) \Big) D_0^{-1} (\lambda_k) = 0 \ , 
\\[3ex] \dis
D_0 (\mu_k) \Big( - \ket{b_k},_\nu \bra{q_k} - \ket{b_k} \bra{q_k},_\nu +  
\ket{b_k} \bra{q_k} U_\nu (\mu_k) \Big) = 0 \ ,
\ea \ee
where $k=1,\ldots,N$ and $\nu = 1,\ldots,m$. Moreover,
\be
 D_0 (\mu_k) = \left( I + \sum_{j=1}^N  \frac{ \ket{s_j}\bra{a_j} }{\mu_k - \lambda_j}\right) \ , 
\quad 
D_0^{-1} (\lambda_k) = \left( I + \sum_{j=1}^N \frac{ \ket{b_j}\bra{q_j} }{\lambda_k - \mu_j} \right) \ .
\ee

\begin{lem} \label{rank}
If $\ket{a}$ and $\bra{b}$ have the maximal rank, then:
\be  \label{conseq}
  \ket{a} \bra{b} = 0 \quad \Longleftrightarrow \quad \ket{a} = 0 \quad  {\rm or} 
\quad \bra{b} = 0 \ . 
\ee 
\end{lem}

\no Proof: immediately follows from the definition of the maximal rank. All columns of $\ket{a}$ (and all rows of $\bra{b}$) have to be linearly indpendent. 
\hfill $\Box$\ods

Using \rf{DD-1-Dir} and applying Lemma~\ref{rank} to equations \rf{DD-2-Dir}, we get the following linear system:
\be
 \bra{a_k},_\nu = - \bra{a_k} U_\nu (\lambda_k) \ , \qquad \ket{b_k},_\nu = U_\nu (\mu_k) \ket{b_k} \ , 
\ee
which is satisfied by:
\be  \label{akbk}
\bra{a_k} = \bra{a_{k0}} \Psi^{-1} (\lambda_k) \ , \qquad 
\ket{b_k} = \Psi (\mu_k) \ket{b_{k0}} \ , 
\ee
where $\bra{a_{k0}}$ and $ \ket{b_{k0}}$ are constant. If $\Psi$ is regular at 
$\lambda_k$ and $\mu_k$, then the solution given by \rf{akbk} 
is general (compare Section~\ref{general}). 

\subsection{Symmetric representation of the Darboux matrix}
\label{symm}

We proceed to derive compact formulas for the remaining ingredients of $D$, namely for $\bra{q_k}$ and $\ket{p_k}$. 
Taking into account Lemma~\ref{rank} we can simplify 
equations \rf{DD-1-Dir}:  
\be  \ba{l}  \dis  \label{DD1}
\bra{a_k} + \sum_{j=1}^N M_{kj} \bra{q_j} = 0 \ ,  \qquad 
\ket{b_k} - \sum_{j=1}^N \ket{s_j} M_{jk} = 0 \ ,  \\[3ex] \dis
\bra{q_k} + \sum_{j=1}^N K_{kj} \bra{a_k} = 0 \ , \qquad 
\ket{s_k} - \sum_{j=1}^N \ket{b_j} K_{jk} = 0 \ ,
\ea \ee
where
\be  \label{MK}
  M_{kj} = \frac{ \scal{a_k}{b_j} }{\lambda_k - \mu_j } \ , \qquad
 K_{jk} = \frac{ \scal{q_j}{s_k} }{\mu_j - \lambda_k} \ .
\ee 
The expression $\scal{a_k}{b_j}$ denotes matrix multiplication: $\scal{a_k}{b_j} = \bra{a_k} \ket{b_j}$ (for any fixed $j, k$). The resulting matrix is not necessarily quadratic. The number of its columns is ${\rm rk} (B_j)$ and the number of its rows is ${\rm rk} (A_k)$. Similarly,  
$\scal{q_j}{s_k}$ is also a matrix (for any fixed $j,k$).  

\begin{rem}
Matrices $M_{jk}$ form the so called ``soliton correlation matrix'' $\hat M$ which has $\sum_{j=1}^N {\rm rk} (B_j)$ columns and $\sum_{k=1}^N {\rm rk}  (A_k)$ rows. 
From \rf{DD1} it follows that 
\be  \label{Minv}
\hat K = {\hat M}^{-1} \ . 
\ee
Therefore, $\hat M$ and $\hat K$ have to be quadratic matrices, i.e., 
\be
\sum_{k=1}^N {\rm rk}  (A_k) = \sum_{j=1}^N {\rm rk} (B_j) \ .  
\ee
\end{rem}

The soliton correlation matrix $\hat M$ is a Cauchy-like matrix (compare \cite{NAH}) which has been reobtained 
several times in various particular cases (see, for instance, \cite{Ci-Nsol,HSA3,Sa-IP,TU}).

\begin{cor}  \label{cor-D-sym}
The symmetric form of the multipole Darboux matrix is given by 
\be  \ba{l} \label{D-sym}  \dis
D (\lambda) = \N \left(   I + \sum_{k=1}^N \sum_{j=1}^N \frac{ \ket{ b_j } \ 
K_{jk} \ \bra{ a_k  } }{\la-\la_k} \right) \ ,  \\[4ex]
\dis
D^{-1} (\lambda) = \left(   I - \sum_{k=1}^N \sum_{j=1}^N \frac{ \ket{ b_j } \ 
K_{jk} \ \bra{ a_k  } }{\la-\mu_j} \right) \N^{-1} \ ,
\ea \ee
where $\N$ is a normalization matrix (we assume  
$\det\N\neq 0$), \ $\hat K = \hat M^{-1}$, \  $\hat M$ is given by \rf{MK}, and $\ket{b_j}$, $\bra{a_j}$ ($j=1,\ldots,N$) are expressed by \rf{akbk}. 
\end{cor}

\subsection{How to represent $N$-soliton surfaces?}

Iterated Darboux matrix is a composition of $N$ binary Darboux transformations (see, for instance, \cite{LRS,ZMNP}):
\be  \label{iter}
D = \N \left( I + \frac{\lambda_N - \mu_N}{\lambda - \lambda_N} P_N \right) 
\ldots \left( I + \frac{\lambda_2 - \mu_2}{\lambda - \lambda_2} P_2 \right)
\left( I + \frac{\lambda_1 - \mu_1}{\lambda - \lambda_1} P_1 \right) \ , 
\ee
where projectors $P_k$ are defined by
\be
 \ker P_k = \Psi_{k-1} (\lambda_k) \ , \quad {\rm im} P_k = \Psi_{k-1} (\mu_k) \ ,
\ee
where $\Psi_k$ are defined by: $\Psi_0 (x,\lambda) = \Psi (x, \lambda)$ and (for $k \geqslant 1$):
\be
  \Psi_k (\lambda) := \left( I + \frac{\lambda_k - \mu_k}{\lambda - \lambda_k} P_k \right) \Psi_{k-1} (\lambda) \ .
\ee
In this case \rf{DBT-F} yields
\be  \label{Fiter}
 \tilde F = F + \sum_{k=1}^N  \frac{ (\mu_k - \lambda_k) \lambda,_\zeta }{(\lambda-\lambda_k)(\lambda  
- \mu_k)} \Psi_{k-1}^{-1} (\lambda) \ P_k \ \Psi_{k-1} (\lambda) 
\ee
Note that the formula \rf{Fiter}  does not contain $\N$. Indeed, gauge equivalent linear problems have identical soliton surfaces, see \cite{Sy6}.

\begin{rem} The determinant of the iterated Darboux matrix \rf{iter} can be easily computed 
(compare \cite{Ci-dbt}): 
\be
  \det D  = \det \N   \prod_{k=1}^N \left( \frac{\lambda-\lambda_k}{\lambda-\mu_k } \right)^{\dim{\rm im} P_k} 
\ee

\end{rem}

Usually the formula \rf{Fiter} is used in the isospectral $SU(n)$ case, when 
$\lambda,_\zeta \equiv 1$,   $\mu_k = \bar \lambda_k$ (see Section~\ref{redu}),  and $\Tr F = \Tr \tilde F = 0$ (what  can be attained by multiplying $D$ by an  appropriate factor $f$, see \rf{F+f}):
\be  \label{Fsun}
   \tilde F = F + \sum_{k=1}^N  \frac{ 2 {\rm Im} \lambda_k}{ |\lambda - \lambda_k|} 
\Psi_{k-1}^{-1} (\lambda) \ i \left( \frac{\dim {\rm im} P_k }{n} \ I - P_k \right) \Psi_{k-1} (\lambda) 
\ , \ee
(in this case projectors are orthogonal, $P_k^\dagger = P_k$, see for example \cite{Ci-dbt,LRB,Sym}). The sum on the right-hand side of \rf{Fsun} consists of traceless components of constant length (using the Killing-Cartan form $a\cdot b = - n \Tr (a b)$ as a scalar product in $su(n)$), see \cite{Sym}. Thus the formula \rf{Fsun} generalizes the classical Bianchi-Lie transformation for pseudospherical surfaces \cite{Sy2,Sym}.

The formula \rf{Fiter} is not manifestly symmetric with respect to permutations 
of $\lambda_k$. 
The symmetric formula for the Darboux-B\"acklund transformation for soliton surfaces can be obtained by substituting \rf{D-sym} into \rf{DBT-F}. 

\begin{Th}  \label{Th-solsur}
The symmetric representation for $N$-soliton surfaces has the form:
\be  \label{F-sym}
 \tilde F = F - \lambda,_\zeta \sum_{j= 1}^N \sum_{k= 1}^N \frac{\Psi^{-1} (\lambda) \ket{b_j}\ K_{jk}\ 
\bra{a_k} \Psi(\lambda)}{(\lambda-\lambda_k)(\lambda-\mu_j)} \ ,  
\ee
(the notation is explained in Corollary~\ref{cor-D-sym}, see also Section~\ref{sec-Sym}).
\end{Th}

\no Proof. We compute $D^{-1} D,_\lambda$ where $D$ is given by \rf{D-sym}:
\[
- D^{-1} D,_\lambda = \sum_{j,k=1}^N \frac{ \ket{b_j} K_{jk} \bra{a_k} }{(\lambda-\lambda_k)^2 } - \sum_{i,j,k,l=1}^N \frac{ \ket{b_j} K_{ji} \scal{a_i}{b_l} K_{lk} \bra{a_k} }{ (\lambda-\lambda_k)^2 (\lambda-\mu_j) } \ .
\]
We use \rf{MK} and perform the summation over $i, l$ in the second component:
\be  \label{uzel}
\sum_{i,l=1}^N K_{ji} (\lambda_i - \mu_l) M_{il} K_{lk}  = 
\sum_{i=1}^N K_{ji} \lambda_i \hat \delta^A_{ik} - \sum_{l=1}^N \hat \delta^B_{jl} \mu_l K_{lk} 
= K_{jk} (\lambda_k-\mu_j) ,
\ee
where $\hat \delta^A_{ik}$, $\hat \delta^B_{jl}$ are natural generalizations of Kronecker's delta (e.g., $\delta^A_{kk}$ is unit matrix of order ${\rm rk} (A_k)$ and $\delta^B_{jj}$ is unit matrix of order  ${\rm rk} (B_j)$). Finally, by virtue of an obvious identity $(\lambda-\mu_j) - (\lambda_k- \mu_j) = \lambda-\lambda_k$, we get \rf{F-sym}. 
\hfill $\Box$\ods

 The expression \rf{F-sym} is a generalization of the symmetric formulas for $N$-soliton surfaces which has been earlier obtained in the $su(2)$-AKNS case (\cite{Ci-Nsol}, see also \cite{Ci-PWN}).

\ods

\begin{prop}  \label{bezsla} 
The Darboux matrix $f D$, with $D$ given by \rf{D-sym} and $f$ given by
\be  \label{f-sym}
f = \sqrt[n]{\prod_{k=1}^N \frac{(\lambda-\lambda_k)^{{\rm rk} A_k} }{(\lambda-\mu_k)^{{\rm rk} B_k} } } 
\ee
transforms traceless $F$ into traceless $\tilde F$. What is more, $\det (f D) = \det N$.
\end{prop}
\ods

\no  Proof: If $D$ is given by \rf{D-sym}, then, using \rf{F-sym}, we compute: 
\be \label{trace}
\frac{\Tr (F - \tilde F )}{ \lambda,_\zeta} =  \Tr \left( \sum_{j,k=1}^N \frac{\scal{a_k}{b_j} 
K_{jk} }{(\lambda - \lambda_k)(\lambda-\mu_j)} \right) = 
\sum_{k=1}^N  \left(  \frac{ \ {\rm rk} (A_k)}{\lambda-\lambda_k}   - \frac{  \ {\rm rk} (B_k) }{\lambda-\mu_k} \right) ,
\ee
where we took into account that $\sum_{j=1}^N M_{kj} K_{jk}$ is the unit matrix of order  ${\rm rk} (A_k)$ and $\sum_{k=1}^N K_{jk} M_{kj}$ is the unit matrix of order  ${\rm rk} (B_j)$.  
Multiplying $D$ by a $\lambda$-dependent function we can change $\tilde F - F$, 
by virtue of \rf{F+f}.  In order to get $\Tr \tilde F = \Tr F$, 
  we have to take  $f$ such that the right-hand side of \rf{trace} equals 
 $n (\ln f),_\lambda$. Hence we get \rf{f-sym}. 

Surprisingly enough, in this way we can compute also the determinant of $D$ given by \rf{D-sym}. Indeed, from \rf{DBT-F} we have $\Tr ( (f D)^{-1} (f D),_\zeta) =0$ (provided that $\Tr \tilde F = \Tr F$) and then Theorem~\ref{Liouville} implies that $\det (f D)$ does not depend on $\zeta$ (and is $\lambda$-independent, as well). Therefore we can evaluate $\det (f D)$  at $\lambda=\infty$. Thus we obtain $\det (f D) = \det \N$.   \hfill $\Box$ 

\ods

Let $\tilde \Psi = f D \Psi$, where $D$ is given by \rf{D-sym} and $f$ is given by \rf{f-sym}. 
For simplicity we assume also \ ${\rm rk} (A_k) = {\rm rk} (B_k) = r_k$. Then:
\be \label{F-SL}
\tilde F = F + \lambda,_\zeta   \Psi^{-1}(\lambda) \sum_{j,k=1}^N  \left( \frac{ (\lambda_k - \mu_k) r_k \delta_{jk} - \ket{b_j} K_{jk} \bra{a_k} }{(\lambda - \lambda_k)(\lambda-\mu_j)} \right) \Psi (\lambda) \ ,
\ee 
where $\delta_{jk}$ is Kronecker's delta. 
Note that the symmetric form of \rf{Fsun} is given by the specialization of the 
formula \rf{F-SL} to the case $\mu_k = \bar\lambda_k$. 

\ods

Another representation for multisoliton surfaces can be derived from the polynomial representation of $D$. 
In order to compute $N$-soliton addition to the  surface 
$F:=\Psi^{-1} \Psi,_\lambda |_{\lambda=\lambda_0}$  we assume the Darboux matrix in 
a general form 
\[
   D = \sum_{k=0}^N T_k (\lambda-\lambda_0)^k \ .
\] \par \no 
The matrices $T_1,\ldots,T_N$ are computed from the following linear system:
\[
    \sum_{k=0}^N T_k (\lambda_\nu - \lambda_0)^k \Psi(\lambda_\nu) p_\nu = 0 \ ,
    \quad (\nu=1,\ldots,Nn) \ ,
\] \par \no 
where $\lambda_\nu \in {\bf C}$ and $p_\nu \in {\bf C}^n$ are constant, and $T_0$ is a given normalization matrix.
Of course, one should take care of reductions what  can result in
some constraints on $\lambda_\nu$, $p_\nu$ and also on $T_0$, see Section~\ref{redu}.
The formula \rf{DBT-F} assumes the form: 
\be
\tilde F =  F +  \Psi^{-1} (\lambda) \  T_0^{-1} T_1 \ \Psi (\lambda) \equiv F + \Psi^{-1} (\lambda) \  \theta_1 \ \Psi (\lambda) \ , 
\ee
where $\theta_1$ is given by \rf{Ck}. Note that also in this case $\tilde F$ does not depend on the normalization matrix $T_0$ (a change of $T_0$ implies such change of $T_1$ that $\theta_1$ 
remains unchanged, compare Section~\ref{ex-multi}).

\section{Group reductions}
\label{redu}

The so called reduction group was introduced by Mikhailov \cite{Mi-group} 
and detailed description of various reductions is given in \cite{Mi,ZM-CMP}, see also \cite{Ci-dbt}. Group reductions (under a different name) found a rigorous treatment in the framework of the loop group theory  
\cite{Gue,PS-loop,TU}, compare Section~\ref{sec-loop}. 

In this section we describe several important types of reduction groups. 
We consider only the case of non-degenerate normalization $\det\N \neq 0$ (which means that $\det \hat D (\lambda)$ is a polynomial of degree $N n$).  
The most convenient form of the Darboux matrix depends on the reduction. 
The polynomial form \rf{Dpol} is very good for reductions to twisted groups, 
the fractional form \rf{D-frac} (and especially its symmetric version \rf{D-sym}) is appropriate for unitary and orthogonal reductions. 
 Then, as an example, we present in more detail the principal chiral model (sigma model) and its reductions. The symmetric form \rf{D-sym} is of great advantage in this case. 

\subsection{Reductions to twisted loop groups}
\label{sec-twist}

Twisted loop groups are defined by $\Psi (\omega \lambda) = Q \Psi (\lambda) Q^{-1}$ where 
$\omega = \exp \frac{2\pi i}{K} $  (hence $\omega^K = 1$) and, necessarily, $Q^K = I$ (we assume also 
$Q = \const$), compare \cite{Gue}. An important example, two dimensional Toda chain (then $K = n$), is discussed in detail in \cite{Mi}, using the fractional representation of $D$ \rf{D-frac}. Here we present a different approach, based on the polynomial representation.

Usually it is better to consider some natural extensions of loop groups which 
follow from the form of the linear problem. The starting point is the assumption about the form of the linear problem (i.e., $U_\nu$ are constrained to the corresponding Lie algebra): 
\be  \label{U-twist}
  U_\nu (\omega \lambda) = Q U_\nu (\lambda) Q^{-1} \ ,
\ee
which implies $( \Psi (\omega\lambda)),_\nu = U_\nu (\omega\lambda) \Psi (\omega\lambda) = 
Q U_\nu (\lambda) Q^{-1} \Psi (\omega\lambda)$. Hence
\[
 \left(  Q^{-1} \Psi (\omega\lambda) \right),_\nu  = U_\nu (\lambda) \left( Q^{-1} \Psi (\omega\lambda) \right) \ , 
\]
which means (see Remark~\ref{rem-C}) that $Q^{-1} \Psi (\omega\lambda) = \Psi (\lambda) C_0 (\lambda)$, where the matrix $C_0 (\lambda)$ does not depend on $x$. Therefore
\be
  \Psi (\omega \lambda) = Q \Psi (\lambda) C_0 (\lambda) \ , 
\ee
and similar equation for $\tilde \Psi = \hat D \Psi$ (with a different $\tilde C_0 (\lambda)$, in general). Therefore: $\hat D(\omega \lambda) Q \Psi (\lambda) C_0 (\lambda) = Q \hat D (\lambda) \Psi (\lambda) \tilde C_0 (\lambda)$. In order to eliminate $\Psi (\lambda)$ we have to assume 
that $\tilde C_0 (\lambda) = \gamma_0 (\lambda) C_0 (\lambda)$, where $\gamma_0: \lambda \rightarrow \gamma_0 (\lambda) \in \C$ is a rational complex function of $\lambda$. Then
\be  \ba{l}
  \hat D (\omega\lambda) = \gamma_0 (\lambda) Q \hat D (\lambda) Q^{-1} \ , \\[2ex]
\det \hat D (\omega\lambda) = (\gamma_0 (\lambda))^n \det \hat D (\lambda) \ .
\ea \ee

\begin{rem}
Computing $\hat D (\omega^2 \lambda)$,\ldots,$\hat D (\omega^K \lambda)$ we obtain a necessary constraint for $\gamma_0$: 
\be  \label{cons-omega}
\gamma_0(\lambda) \gamma_0(\omega\lambda) \ldots \gamma_0 (\omega^{K-1} \lambda) = 1 \ . 
\ee
This constraint is satisfied by any meromorphic function such that  $\gamma_0 (\infty) = 1$ 
and all its zeros and poles coincide with some zeros of $\det\hat D(\lambda)$.  Note that the matrix $C_0 (\lambda)$ also is not arbitrary but satisfies an analogical constraint.
\end{rem}

We make usual assumptions: $\gamma_0 (\lambda) \equiv  1$  and $C_0 (\lambda) \equiv Q^{-1}$ (then $\Psi$ and $\hat D$ are fixed points of the reduction group \cite{Mi}, or, in other words, $\hat D$ and $\Psi$ take values in the loop group). Then
\be  \label{om-loop}
   \Psi (\omega^k \lambda) = Q^k \Psi (\lambda) Q^{-k} \ , \quad 
\hat D (\omega^k \lambda) = Q^k \hat D (\lambda) Q^{-k} \ ,
\ee
for $k=1,\ldots,K-1$. 

\begin{lem}
Let \ $\gamma_0 (\lambda) \equiv 1$. If \ $\det \hat D(\lambda_1) = 0$, then \ $\det \hat D(\omega^k \lambda_1) = 0$  for $k=1,\ldots,K$.  Multiplicities of all these $K$ zeros are  identical. 
\end{lem}

Therefore, if $\hat D(\lambda)$ is a polynomial of order $N$ (and, as a consequence, $\det \hat D (\lambda)$ has the order $N n$), then $N n$ has to be divided by $K$, i.e., there exists an integer $\hat N$ such that  $N n = \hat N K$ (in the two-dimensional Toda chain case $\hat N = N$). Moreover, \rf{om-loop} (evaluated at $k=1$) imply that 
\be   \label{om-T0}
\omega^N T_0 Q = Q T_0 \ ,  
\ee
where $T_0$ is the normalization matrix, compare \rf{Dpol}. We have to demand that this equation 
has a solution $T_0 \neq 0$ (otherwise, the Darboux matrix cannot be a polynomial of order $N$). 
Certainly \rf{om-T0} has a solution for $N$ such that $\omega^N = 1$ (in fact, this assumption was  done in \cite{Mi}).  

\begin{cor}  \label{cor-zer1}
If $\gamma_0 (\lambda) \equiv 1$, then the set of zeros of \ $\det \hat D (\lambda)$ is given by \ $\{ \omega^k \lambda_j |\ k=0,1,\ldots,K-1; \  j = 1,\ldots,\hat N \}$,  where $\lambda_j \in \C$.  
\end{cor}

Equations \rf{TFi}, defining the Darboux matrix, can be rewritten as follows (taking into account  \rf{om-loop}):
\be
0 = \hat D (\omega^k \lambda_j) \Psi (\omega^k \lambda_j) p_{jk} = Q^k \hat D (\lambda_j) \Psi (\lambda_j) Q^{-k} p_{jk} 
\ee
For simplicity we assume the generic case, i.e., all zeros $\omega^k \lambda_j$ are pairwise different (and, as a consequence, simple). Then the kernels of $\hat D (\lambda_k)$ are one-dimensional, which means that $\Psi (\lambda_j) Q^{-k} p_{jk}$ is proportional to $\Psi (\lambda_j) p_{j0}$.  

\begin{Th}
Assuming that \rf{om-T0} has a solution $T_0 \neq 0$ we  
construct the Darboux matrix (a $\lambda$-polynomial of order $N$) according to Corollary~\ref{cor-Dar} taking into account  that its zeros are given by $\omega^k \lambda_j$ (see Corollary~\ref{cor-zer1}) and the corresponding eigenvectors are related by $p_{jk} = Q^k p_{j0}$ ($k=0,1,\ldots,K-1$; \  $j = 1,\ldots,\hat N $, where $\hat N = N n /K$). This Darboux matrix preserves twisted loop group constraints \rf{U-twist} and \rf{om-loop}, i.e., $\tilde U_\nu (\omega\lambda) = Q \tilde U_\nu (\lambda) Q^{-1}$, etc.  
\end{Th}

\begin{rem}  \label{nonis-twist}
In the nonisospectral case twisted reductions impose constraints on the form of $L_\nu$. \
If \rf{lambdak} are satisfied, then  also $\omega \lambda_k,_\nu = L_\nu (x, \omega \lambda_k)$. 
Hence we get the constraint: $\omega L_\nu (x, \lambda) = L_\nu (x, \omega \lambda)$. 
\end{rem}

\ods

The particular case $K=2$ (i.e., $\omega = -1$)  is very popular (e.g., this reduction 
is necessary to derive the standard linear problem for the famous sine-Gordon equation 
\cite{ZMNP,RS-dbt}, see also \cite{Ci-hyper}). 
This case can be generalized by admitting a $\lambda$-dependence of $Q$ (actually such generalization can be done for any $K$ but the results have more complicated form, so we omit them). 
One can easily see that $Q=Q(\lambda)$ has to satisfy 
\be 
Q(-\lambda) Q(\lambda) = \vartheta_0 (\lambda) I \ ,
\ee
where $\vartheta_0 $ is a scalar function such that $\vartheta_0 (-\lambda) = \vartheta_0  (\lambda)$ 
(in particular, we can take $\vartheta_0  (\lambda) \equiv 1$). Assuming $\gamma_0 =1$ we have 
$\det {\hat D} (-\lambda) = \det {\hat D} (\lambda)$ which means that zeros of $\det {\hat D} (\lambda)$ appear in pairs $\lambda_{k'} = - \lambda_k$. 
Constant eigenvectors $p_{k'}$ and $p_k$ satisfy
$\tilde{\Psi}(\lambda_k) p_k = 0$ and $\tilde{\Psi}(\lambda_{k'}) p_{k'} = 0$ which implies
$\tilde{\Psi}(\lambda_k) Q(-\lambda_k) p_{k'} = 0$. If the zero $\lambda_k$ is simple
then $p_{k'} = Q(\lambda_k) p_k$ (the eigenvectors are defined up to a scalar constant 
factor, therefore we omitted the factor $\vartheta_0  (\lambda_k)$). Moreover, the condition \rf{om-T0} should be replaced by $\omega^N T_0 Q_\infty = Q_\infty T_0$, where $Q_\infty$ is either $Q(\infty)$ or the coefficient by the highest power of $\lambda$ in the asymptotic expansion of $Q (\lambda)$ for $\lambda \rightarrow \infty$.

\subsection{Reality condition}
\label{sec-reality}

The condition
\be
   \overline{ U_\nu (\lambda) } = U_\nu (\bar\lambda) 
\ee
(where the bars denote complex conjugates) simply means that all coefficients of matrices $U_\nu$ are real. Considering \rf{ZS} we get
\be  \label{realpsi}
\overline{ \Psi (\bar\lambda) } = \Psi (\lambda) C (\lambda)
\ee
where $\overline{ C (\lambda) } C (\bar\lambda) = I$. Applying \rf{realpsi} to $\tilde \Psi (\lambda) = \hat D (\lambda) \Psi (\lambda)$ we obtain  the corresponding constraint on $\hat D$
  \be
    \overline{ \hat D (\bar\lambda) } = \gamma (\lambda) \hat D (\lambda) \ , 
\ee
where $\gamma (\lambda)$ is a scalar rational function which satisfies $\overline{\gamma (\lambda)} \gamma (\bar\lambda) = 1$. Hence
\be
 (\gamma (\lambda) )^n = \frac{ \overline{ \det \hat D (\bar\lambda) } }{ \det \hat D (\lambda) } \ ,
\ee
which means that $\gamma =  \overline{ w (\bar\lambda) }/w (\lambda)$, 
where $w (\lambda)$ is a polynomial of degree $K \leqslant N$ (provided that $\hat D$ is a polynomial of degree $N$). Then $\det D (\lambda)$ has $K$ zeros of multiplicity $n$ and $(N-K) n$ simple zeros. The set of simple zeros is invariant with respect to the complex conjugation, i.e., they are either real or form pairs of conjugate numbers. 

We assume the simplest case $\gamma (\lambda) \equiv 1$ (and also $C (\lambda) = \tilde C (\lambda) \equiv 1$). Then all zeros of $\det \hat D (\lambda)$ are simple, and either $\lambda_j \in \R$ (then $\bar p_j = p_j$) or there are pairs $\lambda_{k'} = \bar \lambda_k$  (then 
$p_{k'} = \bar p_k$).

\subsection{Unitary reductions}

Unitary reductions (which sometimes are also referred to as reality conditions, see for instance \cite{TU}) are defined by
\be  \label{U-reality}
U_{\bar \nu}^{\dagger} (\bar{\lambda}) = - H  U_\nu (\lambda) H^{-1}  \ , 
\ee
where the dagger denotes the Hermitean conjugate, $H$ is a constant Hermitean matrix 
($H^\dagger = H$) and $\bar\nu$ means complex conjugate (necessary if $x^1, x^2$ are complex,  e.g., usually $x^1 = z$, $x^2 = \bar z$ \ in the case of chiral models, discussed in Section~\ref{sec-chiral}). 
 Using \rf{U-reality} we obtain from \rf{ZS}:
\be  \label{U-re1} 
(\Psi^\dagger (\bar\lambda)),_\nu \equiv (\Psi (\bar\lambda),_{\bar\nu} )^\dagger = 
\Psi^\dagger (\bar\lambda) U_{\bar \nu}^{\dagger} (\bar{\lambda}) \equiv 
-  \Psi^\dagger (\bar\lambda) H (\lambda) U_\nu (\lambda) H^{-1} (\lambda)  .
\ee
Taking into account the well known formula for differentiating the inverse matrix (i.e., $(\Psi^{-1}),_\nu = - \Psi^{-1} \Psi,_\nu \Psi^{-1}$) we transform \rf{U-re1} into 
\be \label{ZS-trans1}
H^{-1} ( (\Psi^\dagger (\bar\lambda))^{-1}),_\nu =  U_\nu (\lambda) 
H^{-1} (\Psi^\dagger (\bar\lambda))^{-1} \ , 
\ee
and, comparing \rf{ZS-trans1} with \rf{ZS}, we get  
\be  \label{PsiC}
( \Psi^{\dagger}(\bar{\lambda}) )^{-1} = H \Psi (\lambda) C_0 (\lambda) \ , 
\ee
where $\Psi (\lambda)$ solves the system \rf{ZS} and $C_0 (\lambda)$ is an $x$-independent matrix. 
From \rf{PsiC} we can derive $C_0^\dagger (\bar \lambda) = C_0 (\lambda)$. 
$\tilde{\Psi} (\lambda)$ satisfies the constraint \rf{PsiC} with $\tilde C_0$ in the place of $C_0$. Assuming $C_0 (\lambda) = k_0 (\lambda) \tilde C_0 (\lambda)$, where $k_0 (\lambda)$ is 
an $x$-independent  scalar function, we  derive
the condition 
\be  \label{DH}
 {\hat D}^\dagger (\bar \lambda) = k_0 (\lambda)  H {\hat D}^{-1}(\lambda) H^{-1} \ ,
\ee  \par \no 
which is necessary for ${\hat D} $ to be the
Darboux matrix in the case of unitary reductions. 
We point out that $k_0 (\lambda)$ has to be a rational function. The simplest choice $k_0 (\lambda) \equiv 1$ is not possible. Indeed, from
\rf{DH} we obtain
\be   \label{k^n}
(k_0 (\lambda))^n  = \overline{\det {\hat D} (\bar{\lambda})} \det {\hat D} (\lambda)  \ .
\ee 
Hence $k_0 (\lambda)$ is a polynomial of degree $2N$ (provided that $\det T_0 \neq 0$). 
Moreover, $\overline{ k_0 (\bar\lambda)} = k_0 (\lambda)$ which means that $k_0 (\lambda)$ is a polynomial with real coefficients. The set of its zeros is symmetric with respect to the real axis.  We confine ourselves 
to the case of $k_0 (\lambda)$ without real roots, i.e.
\[
k_0 (\lambda) = |\det T_0|^2 (\lambda-\lambda_1)(\lambda-\bar{\lambda}_1)\ldots(\lambda-\lambda_N)(\lambda-\bar{\lambda}_N)
\] \par \no 
where $N$ is the degree of the polynom ${\hat D} (\lambda)$. Then  
\[
\det {\hat D}  = (\det T_0) (\lambda-\lambda_1)^{n-d_1}(\lambda-\bar{\lambda}_1)^{d_1}\ldots
          (\lambda-\lambda_N)^{n-d_N}(\lambda-\bar{\lambda}_N)^{d_N} \ ,
\] \par \no 
where $d_k$ are some integers, $1 \leqslant d_k \leqslant n-1$. Note that for $n>2$ the zeros of $\det {\hat D} $ 
are, as a rule, degenerated. If $\lambda_k$ are pairwise different, then dividing ${\hat D} $ by
$(\lambda-\lambda_1)\ldots(\lambda-\lambda_N)$ we obtain the matrix $D$ (equivalent to ${\hat D} $) bounded for $\lambda \rightarrow \infty$ ($\lim_{\lambda\rightarrow \infty} D (\lambda) = T_0$), with singularities at $\lambda=\lambda_k$ ($k=1,\ldots, N$). Taking into account $\det T_0 \neq 0$, we get:
\be  \label{D-frac-uni}
D = \frac{{\hat D} (\lambda)}{(\lambda-\lambda_1)\ldots(\lambda-\lambda_N)} = \N 
\left( I + \sum_{k=1}^N 
\frac{A_k}{\lambda-\lambda_k}  \right) \ , 
\ee 
where $\N = T_0$ and $A_k$ are some matrices dependent on $x$. The inverse matrix $D^{-1}$ has poles at $\lambda = \bar\lambda_k$ ($k=1,\ldots, N$). 
Therefore $D$ is exactly of the form \rf{D-frac} with $\mu_k = \bar \lambda_k$.
 Note that \rf{DH} can be rewritten as: 
\be  \label{DH-sym}
  D^{-1} (\lambda) = H^{-1} D^\dagger (\bar\lambda)  H \ .
\ee
In what follows we use natural notation: $\ket{ a^\dagger} := ( \bra{a} )^\dagger$, 
$\bra{ b^\dagger}  := ( \ket{b} )^\dagger $. 

\ods
\begin{Th} \label{Th-unitary}
The Darboux matrix of the form \rf{D-sym}, satisfying the additional constraints
\be \label{unitary} \ba{l}
\mu_k = \bar\lambda_k  , \quad  \ket{b_{k0}} = C_0 (\bar\lambda_k) \ket{a_{k0}^\dagger}  , 
\quad \N^{\dagger} H \N = H , \quad {\rm rk} A_k = {\rm rk} B_k  , 
\ea \ee
($k=1,\ldots,N$), preserves the unitary reduction defined by \rf{U-reality} and \rf{PsiC}. 

\end{Th}\ods

\no Proof:  We will show that the constraints \rf{unitary}, imposed on $D$ given by \rf{D-sym}, 
are sufficient to satisfy the equation \rf{DH-sym}. The condition $\mu_k=\bar\lambda_k$ is already assumed. Equating normalization matrices in \rf{DH-sym} we obtain $\N^{-1} = H^{-1} \N^{\dagger} H$. The most convenient way to proceed further is to use symmetric form of the Darboux matrix \rf{D-sym}. Equating residua at both sides of \rf{DH-sym} we get $B_k = A_k^\dagger$ (compare \rf{D-frac}), i.e.,  
\be  \label{un-res}
\sum_{j=1}^N \ket{b_k} K_{kj} \bra{a_j} \N^{-1} = - \sum_{j=1}^N H^{-1} \ket{a^\dagger} 
K_{jk}^\dagger \bra{b_j^\dagger} \N^\dagger H \ .
\ee
In order to satisfy this equation it is sufficient to require
\be  
\ket{a_{k}^\dagger} = H \ket{b_k} \ , 
\ee
what  implies $\bra{b_k^\dagger} = \bra{a_k} H^{-1}$. Indeed, using \rf{MK} we get $M_{jk}^\dagger  = - M_{kj}$, and, as a consequence,  $K_{jk}^\dagger = - K_{kj}$, compare \rf{Minv}. 
\hfill $\Box$

\begin{rem}  \label{C=Hinv}
Assuming $C_0 (\lambda) = H^{-1}$ we rewrite the constraint \rf{PsiC} 
as \ $\Psi^\dagger (\bar\lambda) H \Psi (\lambda) = H$, i.e., $\Psi (\lambda)$ takes values in the same loop group as $D (\lambda)$, compare \rf{DH-sym}.   This assumption is not very restrictive. 
It is sufficient to impose it on initial data (at $x=x_0$). Then it holds for any $x$.   
\end{rem}

\begin{rem}
In the non-isospectral case the unitary reduction imposes constraints \ $\overline{L_\nu (x, \bar\lambda)} = L_\nu (x, \lambda)$ \ on the evolution of $\lambda$, compare Remark~\ref{nonis-twist}. 
\end{rem}

\begin{rem}  \label{rem-det}
By virtue of Proposition~\ref{bezsla} the Darboux-B\"acklund transformation for the reduction $SU(n)$ is generated by the Darboux matrix $f D$, where $D$ is given by Theorem~\ref{Th-unitary} and $f$ is given by \rf{f-sym} with ${\rm rk} B_k = {\rm rk} A_k$ and $\mu_k = \bar\lambda_k$, 
\end{rem}

We point out that the case of $k_0 (\lambda)$ with real roots is more difficult. 
Usually, the assumption is made that $\lambda_k$ are not real, compare \cite{GHZ,Mi,TU}. 
The case of real $\lambda_1, \mu_1$ is solved and discussed in the case of the binary Darboux matrix ($N=1$), see \cite{Ci-dbt}. By iterations one can obtain more general solutions. However, it would be interesting to obtain a compact form the Darboux matrix 
corresponding to arbitrary set of eigenvalues (symmetric with respect to the real axis).

\subsection{Chiral fields or harmonic maps} 
\label{sec-chiral}

As an illustrative example, we will consider the equation 
\be  \label{chir}
( \Phi,_1 \Phi^{-1} ),_2 + (\Phi,_2 \Phi^{-1} ),_1 = 0 \ ,
\ee
which describes harmonic maps on Lie groups (provided that $\Phi$ assumes 
values in a Lie group $G$) \cite{TU,Uh} or, in physical context, 
principal chiral fields \cite{HSA1,ZM-ZETF}. Adding another constraint, $\Phi^2 = I$, we 
get chiral fields (or sigma models) on symmetric (or  Grassmann) spaces \cite{AP,HSA2,StA,ZM-ZETF}. 

The chiral model \rf{chir} is integrable and the associated isospectral Lax pair 
$\Psi,_\nu = U_\nu \Psi$ is of the form \cite{ZM-ZETF}:
\be  \label{Laxchir}
\Psi,_1 = \frac{A_1}{1 - \lambda} \ \Psi \ , 
\qquad  
\Psi,_2 = \frac{A_2}{1+ \lambda} \ \Psi \ .
\ee
The Lax pairs considered in \cite{GHZ} and \cite{Uh} are equivalent to \rf{Laxchir} modulo simple 
transformations of the parameter $\lambda$. 
It is convenient to denote
\be \label{Psi0}
  \Phi (x) = \Psi (x,0) \ .
\ee
Then $A_\nu = \Phi,_\nu \Phi^{-1} $ or, in other words, 
\be
 U_\nu (\lambda) = \frac{\Phi,_\nu \Phi^{-1} }{1 + (-1)^\nu \lambda} \ , 
\ee
and the compatibility conditions,
\be
A_1,_2 + A_2,_1 = 0 \ , \qquad 
A_1,_2 - A_2,_1 + [A_1, A_2] = 0 \ ,
\ee
rewritten in terms of $\Phi$ become identical with \rf{chir}.  

The Darboux-B\"acklund transformation for $\Phi$ (in the case of 
the principal $GL(n,\C)$ chiral model, where there are no restrictions on $\Phi$ 
except non-degeneracy) 
 is given by
\be
\tilde  \Phi = D (0) \Phi 
\ee
where $D (\lambda)$ is represented, for instance, by the symmetric formula \rf{D-sym}. 

The $U(n)$ reduction is defined by the constraint\ $\Phi^\dagger \Phi = I$  and 
adding $\det \Phi = 1$ we get $SU(n)$ principal sigma model, see for instance  \cite{HSA1,Uh}. 
These constraints are preserved by an appropriately modified Darboux matrix, see 
Theorem~\ref{Th-unitary} and Remark~\ref{rem-det}.

Chiral models on Grassmann spaces can be characterized by the additional  constraint: $\Phi^2 = I$. 
This is a quite non-trivial reduction, worthwhile to be considered in detail. 

\begin{prop}
The Lax pair \rf{Laxchir} satisfies the constraints
\be   \label{red-gras}
  U_\nu (\lambda^{-1}) = \Phi,_\nu \Phi^{-1} + \Phi U_\nu (\lambda) \Phi^{-1} 
\ , \qquad (\nu = 1,2)
\ee
if and only if \ 
$\Phi^2 = \const$. 
\end{prop}

\no Proof is straightforward. We check that \ $\Phi U_\nu \Phi^{-1} = - U_\nu$ \ iff \ $\Phi^2 = \const$. Then, we compute
\[
 U_\nu (\lambda^{-1}) = \frac{\lambda \Phi,_\nu  \Phi^{-1}}{\lambda+(-1)^\nu} = 
\Phi,_\nu \Phi^{-1} - U_\nu (\lambda) = \Phi,_\nu \Phi^{-1} + \Phi U_\nu (\lambda) \Phi^{-1} \ ,
\]
what ends the proof. \hfill $\Box$\ods

 The right-hand side of \rf{red-gras} has the form of a gauge transformation. Hence, we immediately have the following conclusions. 

\begin{cor}
If \ $\Phi^2 = I$, then 
\be
( \Phi^{-1} \Psi (\lambda^{-1}) ),_\nu = U_\nu (\lambda) \Phi^{-1} \Psi (\lambda^{-1}) \ ,
\ee
which means that
\be  \label{psi-red-gras}
\Psi (\lambda^{-1}) = \Phi \Psi (\lambda) S_0 (\lambda)  \ , 
\ee
where $ S_0 (\lambda)$ is a constant matrix. One can check that 
$S_0 (\lambda^{-1}) = S_0^{-1} (\lambda)$. 
\end{cor}

\begin{prop}
The Darboux transformation preserves  \rf{red-gras} if 
\be \label{D-chiral}
D (\lambda^{-1}) = \tilde\Phi D (\lambda) \Phi^{-1} \ , \qquad 
\tilde \Phi = D (0) \Phi \ .
\ee
\end{prop}

\no Proof. The constraint \rf{psi-red-gras} for $\tilde \Psi = D \Psi$ reads 
\[
D (\lambda^{-1}) \Psi (\lambda^{-1}) = \tilde \Phi D (\lambda) \Psi (\lambda) S_0 (\lambda) \ .
\]
Using \rf{psi-red-gras} we obtain \rf{D-chiral}. Finally, we apply \rf{Psi0}. 
\hfill $\Box$\ods

\begin{cor}  \label{inwersja}
The formula \rf{D-chiral} implies that divisor of poles of $D (\lambda^{-1} )$ 
has to be exactly the same as divisor of poles of $D (\lambda)$. Inverting \rf{D-chiral} we get that  divisors of poles of $D^{-1} (\lambda^{-1})$ and $D^{-1} (\lambda)$ also should coincide. Therefore both sets of poles, i.e., 
$\{ \lambda_1,\ldots,\lambda_N \}$ and $\{ \mu_1,\ldots,\mu_N \}$, are invariant with respect to the inversion $\lambda \rightarrow \lambda^{-1}$. 
\end{cor}

\begin{Th}  \label{Th-fikwa}
We assume that poles and zeros ($\lambda_k, \mu_k$) of the  Darboux matrix \rf{D-sym} can be combined in the following pairs 
\be  \label{inwers}
  \lambda_{k'} = \frac{1}{\lambda_k} \ , \quad \mu_{k'} = \frac{1}{\mu_k} \ ,  
\ee 
and $\lambda_k^2 \neq 1$, $\mu_k^2 \neq 1$. We assume also $\N = I$ and        
\be  \label{ini}
 \bra{a_{j'0}} = \bra{a_{j0}} S_0 (\lambda_j) \ , \qquad \ket{b_{j'0}} = 
S_0^{-1} (\mu_j) \ket{b_{j0}} \ .
\ee
Under these assumptions the Darboux matrix \rf{D-sym} 
satisfies \rf{D-chiral} and, moreover, 
\be  \label{abf}
\bra{a_{j'}} = \bra{a_j} \Phi^{-1} \ , \qquad \ket{b_{j'}} = \Phi \ket{b_j} \ . 
\ee
\end{Th}

\no Proof. We are going to verify that assumptions of the theorem imply \rf{D-chiral}. First, we will show that assumptions \rf{ini} imply \rf{abf}. Using 
\rf{akbk}, \rf{inwers} and \rf{ini} we get:
\be \ba{l}
\bra{a_{j'}} = \bra{a_{j'0}} \Psi^{-1} (\lambda_{j'}) = \bra{a_{j0}} S_0^{-1} (\lambda_j) \Psi^{-1} (\lambda_j) \Phi^{-1} = \bra{a_j} \Phi^{-1}  \ ,\\[2ex]
\ket{b_{j'}} = \Psi (\mu_{j'}) \ket{b_{j'0}} = \Phi \Psi (\mu_j) S_0 (\mu_j) 
\ket{b_{j'0}} = \Phi \ket{b_j} \ .
\ea \ee
Then, we compute
\be
M_{j'k'} = \frac{ \scal{a_{j'}}{b_{k'}}}{\lambda_{j'} - \mu_{k'}} = 
\frac{  \scal{a_j}{b_k}  \lambda_j \mu_k }{\mu_k - \lambda_j} = 
- \lambda_j \mu_k M_{jk} \ ,
\ee
and, {\it vice versa}, $M_{jk} = - \lambda_{j'} \mu_{k'} M_{j'k'}$.  Hence,
\be  \label{Kprim}
 K_{k'j'} = - \frac{1}{\mu_k \lambda_j} \ K_{kj} \ , \qquad 
 K_{kj} = - \frac{1}{\mu_{k'} \lambda_{j'} } \ K_{k'j'} \ .
\ee
Assuming $\N =I$ we proceed to compute ingredients of the formula \rf{D-chiral}:
\be  \label{D(0)} 
D (0) = I - \sum_{j,k=1}^N \frac{ \ket{b_j} K_{jk} \bra{a_k}  }{\lambda_k} 
= I + \sum_{j,k=1}^N \frac{ \Phi^{-1} \ket{b_j} K_{jk} \bra{a_k} \Phi  }{\mu_j} \ , 
\ee
where the second equality follows from:
\be
\sum_{j,k=1}^N \frac{ \ket{b_j} K_{jk} \bra{a_k}  }{\lambda_k} = - 
\sum_{j',k'=1}^N  \lambda_{k'} \Phi^{-1} \ket{b_{j'}} \left ( \frac{K_{j'k'}}{\mu_{j'} \lambda_{k'}} \right) \bra{a_{k'}} \Phi  \ ,
\ee
(primes can be dropped because we sum over the same set of  indices). 
Then, we compute $D (\lambda^{-1}) $ and decompose it into the sum of partial fractions:
\be
D (\lambda^{-1}) = I - \sum_{j,k=1}^N \frac{ \ket{b_j} K_{jk} \bra{a_k}  }{\lambda_k} 
- \sum_{j,k=1}^N \frac{ \lambda_k^{-2} \ket{b_j} K_{jk} \bra{a_k}  }{\lambda - \lambda_k^{-1}} \ . 
\ee  
Using \rf{D(0)}, \rf{abf} and \rf{Kprim}, we get (after dropping primes):
\be  \label{ll}
 D (\lambda^{-1}) = D (0)  
+ \sum_{j,k=1}^N \frac{ \lambda_k \Phi^{-1} \ket{b_j} K_{jk} \bra{a_k} \Phi  }{\mu_j (\lambda - \lambda_k) } \ . 
\ee
Finally, 
\be  \label{pp}
\tilde \Phi D (\lambda) \Phi^{-1} =  D(0) + 
\sum_{j,k=1}^N \frac{ \Phi \ket{b_j} K_{jk} \bra{a_k} \Phi^{-1} }{\lambda - \lambda_k} 
+ \sum_{i,j,k,l=1}^N \frac{\Phi^{-1} \ket{b_j} W_{jk} \bra{a_k} \Phi^{-1} }{\mu_j (\lambda - \lambda_k) } 
 \ ,
\ee
where
\be
W_{jk} = \sum_{i,l=1}^N  K_{ji} \bra{a_i} \Phi^2 \ket{b_l} 
K_{lk} =  \sum_{i,l=1}^N  K_{ji} M_{il} (\lambda_i - \mu_l) 
K_{lk} = (\lambda_k - \mu_j) K_{jk}  ,
\ee
where we used $\Phi^2 = I$ and \rf{uzel}. Substituting $W_{jk}$ into \rf{pp} 
and comparing the result with \rf{ll} we get \rf{D-chiral}. \hfill $\Box$\ods

\ods

Usually it is sufficient to assume $C_0 (\lambda) = H^{-1} = \const$ (compare Remark~\ref{C=Hinv}) and $S_0 = \const$ (but the assumption $S_0 = I$ can be too restrictive).

\ods
\begin{prop}  We assume $S_0 = \const$, $H = \const$, $S_0^2 = I$, $H^\dagger=H$ and \ $S_0^\dagger H S_0  = H$. 
We consider the Darboux matrix \rf{D-sym} such that $\N = I$,  $N = 2 K$ and 
\be \ba{l}
\lambda_{j+K} = \lambda_j^{-1} \ , \quad \mu_{j+K} = \mu_j^{-1} \ , \quad \mu_k = \bar\lambda_k \ , \quad |\lambda_j|^2 \neq 0 \ , \quad \bar\lambda_j \neq \lambda_j \ , \\[2ex]
\bra{a_{j0}} = \bra{b_{j0}^\dagger} H \ , \quad  \bra{a_{j+K,0}} = \bra{b_{j+K,0}^\dagger} H,  \\[2ex] 
\bra{a_{j+K,0}} = \bra{a_{j0}} S_0 \ ,  \quad   S_0 \ket{b_{j+K,0} } = \ket{b_{j0} } \ ,
\ea \ee
where $j=1,\ldots,K$, $k=1,\ldots, 2K$. Thus all these data can be expressed by \ 
$\bra{a_{10}},\ldots,\bra{a_{K0}}$ \ and \ $\lambda_1,\ldots,\lambda_K$.  
The Darboux-B\"acklund transformation generated by such Darboux matrix preserves reductions: $\Psi^\dagger (\bar\lambda) H \Psi (\lambda) = H$ and $\Psi (\lambda^{-1}) = \Psi (0) \Psi (\lambda) S_0$. 
\end{prop} 
\ods

\no Proof: 
We apply Theorems~\ref{Th-unitary} and~\ref{Th-fikwa}. It is enough to check whether the equations 
\[
\bra{a_{j0}} = \bra{b_{j0}^\dagger} H, \ \bra{a_{j'0}} = \bra{b_{j'0}^\dagger} H,  \ 
\bra{a_{j'0}} = \bra{a_{j0}} S_0 ,  \  \bra{b_{j0}^\dagger} = \bra{b_{j'0}^\dagger} S_0^\dagger , 
\]
are not contradictory. These equations imply 
$\bra{a_{j0}} H^{-1} = \bra{a_{j0}} S_0 H^{-1} S_0^\dagger$. Hence, using  $S_0^2 = I$, we obtain the constraint  $S_0^\dagger H S_0  = H$  
assuring the compatibility of both reductions. Finally, we denote $j'=j+K$. 
\hfill $\Box$

\section{Connections with other approaches}

In this section we shortly present some other methods of constructing the Darboux-B\"acklund transformation. We show how they are connected with the  approach presented in this paper.

\subsection{Matrix-valued spectral parameter}

The name of Darboux first appeared in the context of the dressing transformations in Matveev's papers (see for instance \cite{Mat}) who extended the notion of 
Darboux covariance, known in the case of the Sturm-Liouville-Schr\"odinger spectral  problems, on arbitrary differential operators \cite{MS}. 

In order to apply Matveev's approach to Zakharov-Shabat spectral 
problems \rf{ZS} the matrix spectral parameter is introduced: 
\be  \label{diag-Lam}
\Lambda := \diag (\lambda_1, \ldots, \lambda_n) \ 
\ee 
(this notation should not be confused with the function $\Lambda$ 
described in Remark~\ref{Lambda}). 
We consider the linear problem of the form \cite{Bob-DB,MS}:
\be  \label{Ma-Bob}
\Psi,_\nu = \sum_j \sum_{k=1}^{N_j} U_{\nu kj} \Psi M_j^k + \sum_{k=0}^{N} V_{\nu k} \Psi
\Lambda^k \ ,
\ee
where $U_{\nu kj}$ and $V_{\nu k}$ are matrices which do not depend on $\lambda_1,\ldots,\lambda_n$ and 
\[
M_j := \diag \left( \frac{1}{\lambda_1 - a_j}, \ldots, \frac{1}{\lambda_n-a_j} \right) \ .
\]
The following theorem holds \cite{Bob-DB,MS}. 

\begin{Th}  \label{Dar-cova}
Equations \rf{Ma-Bob} are covariant with respect to the Darboux transformation
\be  \label{D-Lambda}
		\tilde \Psi = \Psi \Lambda - \sigma \Psi \ , \qquad 
		\sigma = \Psi_1 \Lambda_1 \Psi_1^{-1} \ ,
\ee
where $\Psi_1$ is a fixed solution to \rf{Ma-Bob} with $\Lambda$ replaced by 
the diagonal matrix $\Lambda_1 = \diag (\lambda_{11},\ldots,\lambda_{n1}) $.
\end{Th}

The linear problem \rf{Ma-Bob} is closely related to the following special case of the standard Zakharov-Shabat linear problem \rf{ZS}: 
\be
\Phi,_\nu = \sum_j \sum_{k=1}^{N_j} U_{\nu kj} \frac{1}{(\la-a_j)^k} \Phi  + 
\sum_{k=0}^{N} V_{\nu k} \la^k \Phi \ .
\ee
Namely:
\be      \Psi (\Lambda) = 
\{\Phi(\la_1) p_1,\ldots,\Phi(\la_n) p_n \} \ , 
\ee
where the notation used on the right-hand side (a matrix as a sequence of columns) 
is the same as in \rf{kolumny} and $p_1,\ldots,p_n$ form a constant basis in $\C^n$. 

The Darboux matrix generating the transformation \rf{D-Lambda} 
 can be easily computed (using $D = \tilde \Psi \Psi^{-1}$). We get
\be
D (\Lambda) = \Psi \Lambda \Psi^{-1} - 
\Psi_1 \Lambda_1 \Psi_1^{-1} \ .
\ee 
If we put $\lambda_1 = \ldots = \lambda_n = \lambda$, (i.e., $\Lambda = \lambda I$), and $p_k 
\equiv  e_k$ form the canonical basis in $\C^n$ (i.e., $\{ p_1, \ldots, p_n \} = I$), then $\Phi (\lambda) = \Psi (\lambda I) \equiv \Psi (\lambda)$.   
In this case we obtain 
\be  \label{Gu-form}
  D (\lambda) = \lambda I - \Psi_1 \Lambda_1 \Psi_1^{-1} \ ,
\ee
which is the starting point for the construction of the Darboux matrix 
by Gu and his collaborators \cite{Gu,GHZ,GZ,Zh}. Sometimes another form 
is used:
\be
 D = I - \lambda \Psi_1 \Lambda_1^{-1} \Psi_1^{-1} \ ,
\ee 
which is equivalent to \rf{Gu-form} after changing $\lambda \rightarrow \lambda^{-1}$.

\subsection{Transfer matrix form of the Darboux matrix}

A rational $n\times n$ matrix function $D (\lambda)$, analytic at infinity, can be represented in the form 
\cite{BGK,GKS}:
\be \label{real-D}
  D (\lambda) = \N + F (\lambda I_N  - A)^{-1} G \ , 
\ee
where $A$ is an $N\times N$ matrix, $I_N$ is the unit matrix of order $N$, and $\N, F, G$ 
are matrices of sizes $n\times n$, $n\times N$ and $N\times n$, respectively. Such representation is called a ``realization'' or a ``transfer matrix representation'' of $D$ and the number $N$ (i.e., the order of $A$) is known as the ``state space dimension'' of the realization. 
Realizations are not unique and can have different values of the number $N$.  
``Minimal realizations'' have minimal value of $N$ (and the minimal $N$ 
is called the McMillan degree of $D$). Minimal realizations are unique up to a change of the basis in 
the state space (i.e., $F \rightarrow F T^{-1}$, $A \rightarrow T A T^{-1}$ and $G \rightarrow T G$, \ for some invertible $N\times N$ matrix $T$)  \cite{BGK,GKS}. 
\ods

\begin{prop} 
If \rf{real-D} is a realization for $D$, then one of realizations for $D^{-1}$ is given by
\be  \label{real-invD} 
  D^{-1} (\lambda) = \N^{-1} - \N^{-1} F (\lambda I_N - A + G \N^{-1} F)^{-1} G \N^{-1} \ .
\ee
The realization \rf{real-invD} is minimal iff \rf{real-D} is minimal, see \cite{GKS}. 
\end{prop}

\no The formula \rf{real-invD} can be verified by a simple but non-trivial computation. The obvious identity \ $(\lambda I_N - A + G \N^{-1} F) - (\lambda I_N - A) = G \N^{-1} F$ \ is very helpful, compare \rf{Suzel}. 
\ods

Assuming $\N = I$ we consider the so called transfer matrix 
\be  \label{D-transfer}
     W_A (x,\lambda) = I_n - \Pi_2^* S^{-1} (A - \lambda I_N)^{-1} \Pi_1 \ ,
\ee
where $A, S, \Pi_1, \Pi_2^*$ are some matrices (the star denotes a matrix conjugate, but this is not very important at this moment) and, moreover, the following operator identity holds:
\be  \label{opid}
A S - S B = \Pi_1 \Pi_2^* \ .
\ee
Matrices $A,B,\Pi_1,\Pi_2,S$ satisfying \rf{opid} are said to form an $S$-colligation \cite{Sa-UMN}. 

The transfer matrix \rf{D-transfer} can be used to generate solutions to integrable systems by the Darboux-B\"acklund transformation, see \cite{Sa-IP,Sa-nonis}. 
We can make the following identification:
\be  \ba{l}  \label{coinc}
S = \hat M \ , \quad S^{-1} = \hat K, \\[2ex]
A = \diag (\lambda_1,\lambda_2,\ldots,\lambda_N ) \ , \quad B = \diag (\mu_1,\mu_2,\ldots,\mu_N) 
\ea \ee
 and, finally
\be
\Pi_2^* = \left( \ket{b_1}, \ket{b_2},\ldots, \ket{b_N} \right) \ , \quad 
  \Pi_1 = \m \bra{a_1} \\ \bra{a_2} \\ \vdots \\ \bra{a_N} \ema \ . 
\ee

\begin{cor}
The symmetric representation of the Darboux matrix \rf{D-sym} can be identified with 
the transfer matrix form \rf{D-transfer}, where $A$ is diagonal.  The identity \rf{opid} coincide with the definition \rf{MK} of the matrix $\hat M$.
\end{cor}

Constant matrices $A$ of 
more general form correspond to generalizations of \rf{D-frac} (multiple poles are allowed).  

In order to show a flavour of the transfer matrix technique we present one of typical results. 
Note that the proof of Proposition~\ref{prop-Sakh} is similar to some steps in the proof of  Theorem~\ref{Th-solsur}.

\begin{prop} \label{prop-Sakh} 
We assume the identity \rf{opid} and $D \equiv w_A$ is given by the formula \rf{D-transfer}. 
Then
\be
 D^{-1} = I_n + \Pi_2^* (B - \lambda I_N)^{-1} S^{-1} \Pi_1 
\ee
\end{prop}

\no Proof: We compute
\[
\left( I_n - \Pi_2^* S^{-1} (A - \lambda I_N)^{-1} \Pi_1 \right) 
\left( I_n + \Pi_2^* (B - \lambda I_N)^{-1} S^{-1} \Pi_1  \right) = I_n + \Pi_2^* X \Pi_1 \ , 
\]
where $X$ is $N\times N$ matrix given by
\[
X=( B - \lambda)^{-1} S^{-1} - S^{-1} (A - \lambda)^{-1} - S^{-1} (A-\lambda)^{-1} (A S - S B) (B - \lambda)^{-1} S^{-1} \ .
\]
Now, using the obvious identity 
\be  \label{Suzel}
A S - S B = (A - \lambda) S - S (B - \lambda) \ ,
\ee
we decompose the last component of $X$ into the sum of two terms which immediately cancel with the first two components of $X$. Therefore $X= 0$ which ends the proof. 
\hfill $\Box$ \ods

Vectorial Darboux transformations constitiue one more approach to Darboux transformations,  applied mostly in $2+1$-dimensional case 
\cite{LM,Man}. Although this technique needs no analogue of the Darboux matrix but 
the Darboux transformation is expressed by a Cauchy-like matrix and important role is played by 
operator identities like \rf{opid}. Comparing the results of \cite{Man} and \cite{Sa-IP}  we conclude that both methods are in a very close correspondence (note that the matrix $S$ of \cite{Sa-IP} corresponds to the matrix $\Phi$ of \cite{Man}).

\subsection{Factorization in loop groups}
\label{sec-loop}

Given a Lie group $G$ we define the loop group of $G$ as the group of smooth functions $\gamma : S^1 \rightarrow G$, where $S^1$ denotes the unit circle 
on the complex plane ($|\lambda| = 1$) \cite{Gue,PS-loop}. 
An important role in the loop group theory plays the Birkhoff factorization theorem. The Birkhoff decomposition is closely related to the Riemann-Hilbert problem which provides a rigorous 
background for the inverse scattering method \cite {ZMNP}, see also \cite{Gue}. 

In general the Birkhoff factorization is not explicit. The explicit cases are closely related to the construction of Darboux matrices \cite{Te,TU,Uh} (and also to the construction of finite gap solutions), compare similar ideas in the soliton theory \cite{Its,Kr}. The approach based on the so called cc-ideals is one more link between the loop group theory and the theory of solitons \cite{Har,HS}. 

\ods

From geometrical point of view the Lax pair consists of commuting differential operators and their compatibility can be interpreted as the condition that a one-parameter family of connections is flat:
\be
  \left[ \partial_1 - U_1 (x, \lambda), \ \partial_2 - U_2 (x, \lambda) \right] = 0 
\ee
($U_1, U_2$ are matrices depending on $x$ through some fields, say $u$). 
The ``trivialization'' $E$ of a solution $u$ is defined as the solution of the system:
\be  \label{ZS-E} 
 E,_\nu = - E U_\nu \ , \qquad E (0, \lambda) = I \ .
\ee
Then $E (x,\lambda)$ is holomorphic for $\lambda \in \C$, see \cite{TU}. The function $E (x,\lambda)$ is also referred to as an ``extended solution'', an ``extended frame'' or simply a ``frame''. Comparing \rf{ZS-E} with \rf{ZS} we can identify $E = \Psi^{-1}$. 
Actually, \rf{ZS-E} is the adjoint of \rf{ZS}, see also \rf{ZS-ad}.

\begin{Th} [Birkhoff] 
The multiplication map $\mu$ 
\[
\mu: L_+ ( GL(n,\C)) \times L_- (GL(n,\C)) \rightarrow L (GL(n,\C)) 
\] 
is a diffeomorphism onto an open dense subset of $L (GL(n,\C))$, where  
\begin{itemize}
\item $L_+ ( GL(n,\C))$  is the group of holomorphic maps $h_+ : \C \rightarrow GL(n,\C)$ 
\item $L_- (GL(n,\C))$  is the group of holomorphic maps $h_- : {\cal O}_\infty \rightarrow GL(n,\C)$ such that $h_- (\infty) = I$, where $\cal O_\infty$ is a neighbourhood of $\lambda=\infty$. 
\item $L (GL(n,\C))$ is the group of holomorphic maps from ${\cal O}_\infty \cap \C$ to $GL(n,\C)$ 
\end{itemize}

\end{Th} 

\ods

\begin{cor}
Suppose that $h_- h_+$ lies in the image of $\mu$. Then, by virtue of the Birkhoff theorem, there exists a unique pair $f_\pm \in L_\pm ( GL(n,\C))$ such that 
$h_- h_+ = f_+ f_-$. One can interpret it as a ``dressing action'' of $h_-$ on $h_+$ and $f_+$ is the result of this action, which is denoted by $h_- \sharp h_+ = f_+$.  
\end{cor}

The dressing action seems to ``forget'' about $f_-$. However, it is worthwhile to stress that this is $f_-$ which should be identified with our Darboux matrix. On the other hand the element $h_-$ is deeply hidden (almost non-existing) in other approaches to the construction of Darboux matrices. In order to explain the dressing action generated by the Birkhoff decomposition we will present the binary Darboux transformation \rf{Dbin} 
in the framework of the loop group approach, following \cite{TU}.

We assume that $E(x,\lambda) \in L_+ (GL(n,\C))$ is given, and we choose the so called ``simple element''  $h_{\lambda_1,\mu_1,\pi} \in L_- (GL(n,\C))$:
\be
  h_{\lambda_1,\mu_1,\pi} (\lambda) = I + \frac{\lambda_1 - \mu_1}{\lambda-\lambda_1} \pi \ ,
\ee 
where $\lambda_1, \mu_1$ are complex parameters, and $\pi$ is a constant ($x$-independent) projector in $\C^n$ (i.e., $\pi^2 = \pi$). One can easily see that $h_{\lambda_1,\mu_1,\pi}^{-1} = 
h_{\mu_1,\lambda_1,\pi}$, compare \rf{Dbin} and \rf{Dbin-inv}. 

Then, the Birkhoff theorem states that there exists $\tilde E \in L_+ ( GL(n,\C))$ and $D \in 
L_- ( GL(n,\C))$ such that
\be  \label{Birk1}
    h_{\lambda_1,\mu_1,\pi}  E (x, \lambda) = \tilde E (x,\lambda) D (x,\lambda) \ , 
\ee   
provided that the product on the left-hand side belongs to certain 
``open dense set'' of $L (GL(n,\C))$. Now, both the exact form of $D$ and this ``open dense set'' can be found by direct calculation. It is sufficient (similarly as in all other approaches discussed earlier) to compare the residua at both sides of the equation \rf{Birk1}. Hence
\be
  D = I + \frac{\lambda_1 - \mu_1}{\lambda - \lambda_1} P \ , 
\ee
where $P$ is defined by \rf{14}, where $V_{ker} = \ker \pi$ and $V_{im} = {\rm im}\ \pi$. The open dense set from the Birkhoff theorem is defined by: \  $\ker P \cap {\rm im} P = \{0 \}$. We remark, by the way, that the Birkhoff theorem assumes the isospectral case and the canonical normalization ($\N =I$). 

Note that \rf{Birk1} implies $\tilde E = h E D^{-1} = h \Psi^{-1} D^{-1} = (D \Psi h^{-1} )^{-1} $ (where $h$ denotes the simple element). Therefore, $\tilde \Psi \equiv E^{-1} = D \Psi h^{-1}$, 
which is equivalent (because $h$ does not depend on $x$) to the usual formula $\tilde \Psi = D \Psi$ (compare Remark~\ref{rem-C}).

\section{Invariants of the Darboux transformation}
\label{invariants}

The Darboux transformation changes the matrices $U_\nu$ into new matrices 
$\tilde{U}_k$ of the same form. By invariants of the Darboux transformation we 
mean constraints on coefficients of $U_\nu$ which are preserved by 
the transformation, see \cite{Ci-dbt} (compare also \cite{Sh}, where 
one may find many examples). The invariants are very useful in the construction of Darboux matrices in purely algebraic way, without referring to the special boudary conditions and to the scattering theory (which is a usual practice, compare \cite{GHZ,TU,Zh}).

Here we simplify the approach 
of \cite{Ci-dbt} and extend it on the non-isospectral polynomial case. Moreover, we show that our approach works also in much more general case: when the Lax pair 
is singular at some fixed values of the spectral parameter. In this section we denote $U_1 = U$, $U_2 = V$.

\subsection{Linear invariants for polynomial Lax pairs}

We consider Lax pairs with the following $\lambda$-dependence: 
\be  \label{UV-poly}
U = \sum_{k=0}^{\infty} u_k \lambda^{N-k} \equiv \lambda^N u \ , \qquad V = \sum_{k=0}^{\infty} v_k \lambda^{M-k} \equiv \lambda^M v  \ ,
\ee
where $N, M$ are fixed positive integers (not to be confused with the notation of previous sections) and $u_k = u_k (x)$, $v_k = v_k (x)$ ($k = 0, 1, 2, \ldots$). Usually the sums are finite (i.e., $u_k = v_k =0$ for sufficiently large $k$), and this typical case (polynomial in $\lambda$ and $\lambda^{-1}$) corresponds to many classical soliton equations. In particular both $U$ and $V$ can be polynomials in $\lambda$ (in this case, for $N=1$, we get  famous AKNS hierarchy). We also assume a similar $\lambda$-dependence of the derivatives of $\lambda$:
\be  \label{nonisy}
\lambda,_1 = \sum_{k=0}^\infty a_k \lambda^{N'-k} \ , \qquad \lambda,_2 = \sum_{k=0}^\infty b_k \lambda^{M'-k} \ ,
\ee
where $N', M'$ are given integers fixed by the assumption $a_0 \neq 0$, $b_0 \neq 0$ (in the nonisospectral case). The coefficients $a_k = a_k (x)$, $b_k = b_k (x)$ have to satisfy compatibility conditions resulting from $\lambda,_{12} = \lambda,_{21}$ (some examples can be found in \cite{BZM,Ci-dbt,Sh}). 

We consider the Darboux transformation of $U + H V$, where $H= H (x, \lambda)$  is a fixed function 
\be
  H (x,\lambda) = \lambda^{N-M} h (x,\lambda) \equiv \lambda^{N-M} (h_0 + h_1 \lambda^{-1} + h_2 \lambda^{-2} + \ldots ) \ ,
\ee
where $h_0, h_1, h_2,\ldots$ are  given functions of $x$.  We assume that $H$ is unchanged by the Darboux transformation (and $U, V$ are transformed, as usual, according to \rf{DBT}). The Darboux transformation yields
\be
(\tilde U + H \tilde V) D = D,_1 + h D,_2 + D (U + h V) \ ,
\ee
which reduces to
\be  \label{lin-trans}
(\tilde{u} + h \tilde{v})D - D (u + h v) = 
\lambda^{-N} D,_1 + h \lambda^{-M} D,_2 \ .
\ee
We assume that 
$D$ is analytic at $\lambda = \infty$ :
\be
D = T_0 + T_1 \lambda^{-1} + T_2 \lambda^{-2} + \ldots \ , \qquad  \det T_0 \neq 0 \ ,
\ee
i.e., $D = \lambda^{-N} \hat D$, where $\hat D$ is given by \rf{Dpol}.  
\ods
The idea of linear invariants is quite obvious. Suppose that for 
$\lambda \approx \infty$ the right-hand side of \rf{lin-trans} behaves as $\lambda^{-K}$, where $K \geqslant 1$. Then the first $K$ terms of the Taylor expansion (in $\lambda^{-1}$) of the left-hand side are equal to zero. The first two of these 
equations read:
{\small 
\[ \bal
(\tilde{u}_0 + h_0 \tilde{v}_0) T_0 = T_0 (u_0 + h_0 v_0) \ , \ex
(\tilde{u}_1 + h_0 \tilde{v}_1 + h_1 \tilde{v}_0) T_0 + 
(\tilde{u}_0 + h_0 \tilde{v}_0 ) T_1 =  T_0 ( u_1 + h_0 v_1 + h_1 v_0) + 
T_1 (u_0 + h_0 v_0) \ .
\ea \] }
The asumption $u_0 + h_0 v_0 = 0$ implies $\tilde{u}_0 + h_0 \tilde{v}_0 = 0$ (provided that $\det T_0 \neq 0$). Then, 
adding the second assumption: $u_1 + h_0 v_1 + h_1 v_0 = 0$, we obtain  
as a consequence $\tilde{u}_1 + h_0 \tilde{v}_1 + h_1 \tilde{v}_0 = 0$. 
Thus we have two expressions invariant with respect to the Darboux transformation. Considering the first $k$ (where $k \leqslant K$) equations we 
get an invariant system of $k$ equations.

We proceed to estimate $K$. The leading terms of the right-hand side of \rf{lin-trans} are given by:
\be  \label{lead-1}
\lambda^{-N} (T_0,_1 - a_0 T_1 \lambda^{N'-2} + \ldots) + h_0 \lambda^{-M} 
(T_0,_2 - b_0 T_1 \lambda^{M'-2} + \ldots) 
\ee 
Therefore $K > k_{max1}$, where
\be  \label{kmax1}
k_{max1} =  - 1 + \min \{ N, M, N+2-N', M+2-M' \}   \ ,
\ee
what can be summarized as follows. 

\begin{prop}  \label{lin-prop} 
Suppose that $0 \leqslant k \leqslant k_{max1}$ and $h_0,h_1,\ldots, h_k$ are given functions of $x$. Then the system of\  $k+1$\ linear constraints 
\be  \label{lin-inv}
u_j + \sum_{i=0}^j h_i v_{j-i} = 0 \ ,  \qquad (j=0,1,\ldots,k) \ , 
\ee
is invariant with respect to Darboux transformations such that $\det T_0 \neq 0$. 
\end{prop}

\no  In some special cases, we can formulate stronger propositions (i.e., we have more invariants). In the isospectral case we can replace $k_{max1}$ by 
\be   \label{kmax1-isos}
k'_{max1} =  - 1 + \min \{N, M \} \ ,
\ee
(the same result is valid when $N' \leqslant 2$ and $M' \leqslant 2$).
In the case of the canonical normalization ($T_0 = I$) we can replace $k_{max1}$ 
by
\be   \label{kmax1-canon}
k''_{max1} = \min \{ N, M, N+1-N', M+1-M' \} \ .
\ee
Below we present one more example.

\begin{prop}  \label{prop-g}
Suppose that \ 
$\min \{  M, N+2-N', M+2-M' \} > N$ \ (it implies, in particular, $k_{max1}= N-1$),   
functions $h_0,h_1,\ldots, h_k$ ($k \leqslant N$) are given, and 
$T_0$ assume values in some matrix Lie group $G$. Then, the following system of $k+1$ linear constraints is invariant with respect to the Darboux transformation:
\be \ba{l} \label{lin-inv-Lie} \dis 
u_j + \sum_{i=0}^j h_i v_{j-i} = 0 \ ,  \qquad (j=0,1,\ldots,k-1) \ , \\[3ex] \dis
u_k + \sum_{i=0}^k h_i v_{k-i}  \in    g  \ ,  
\ea \ee
where $g$ is the Lie algebra of the Lie group $G$. 
\end{prop} 

The proof of this proposition is analogical to the proof of Proposition~\ref{lin-prop}: we consider coefficients by powers of $\lambda^{-1}$ in \rf{lin-trans}. 
Only the last step has to be treated in a different way. 
Assuming that the first $k$ constraints hold, the coefficients by $\lambda^{-k}$ yield
\be
\tilde u_k + \sum_{i=0}^k h_i \tilde v_{k-i} = T_0 \left( u_k + \sum_{i=0}^k h_i v_{k-i} \right) T_0^{-1} + \delta_{kN} T_0,_1 T_0^{-1} \ .
\ee
Now the proof follows immediatelly from well known properties of matrix Lie groups ($T g T^{-1} \subset g$ and $T,_1 T^{-1} \in g$, provided that $T = T (x) \in G$).

\subsection{Bilinear invariants for polynomial Lax pairs}

Assuming the polynomial form \rf{UV-poly} of $U, V$ we consider the Darboux transforms of bilinear forms $\Tr (U^2)$, $\Tr (V^2)$ and $\Tr (UV)$.  
We present computations for the last case (the other two cases are analogical). In this setion we use notation: $ A \cdot B \equiv \Tr ( A B)$.  From \rf{DBT} we get
\be  \label{BIL} 
\Tr(\tilde{U} \tilde{V}) -  \Tr(U V)  = \Tr(D,_1 D^{-1} D,_2 D^{-1}  +
D,_1 V D^{-1} +  D,_2 U D^{-1})  \ .
\ee
The leading terms of the right-hand side of \rf{BIL} read
\be \ba{l}  \label{lead-2}
\lambda^{- (N+M)} \Tr \left(  (T_0,_1 - a_0 T_1 
\lambda^{N'-2} + \ldots) T_0^{-1} (T_0,_2 - b_0 T_1 \lambda^{M'-2}+\ldots) T_0^{-1} \right)  ,  \\[3ex]
\lambda^{-N} \Tr \left( (T_0,_1 - a_0 T_1 
\lambda^{N'-2} + \ldots) v_0 T_0^{-1} \right) \ , \\[3ex] 
\lambda^{-M} \Tr \left( (T_0,_2 - b_0 T_1 \lambda^{M'-2}+\ldots) u_0 T_0^{-1} \right) \ . 
\ea \ee
Thus the right-hand side of \rf{BIL} behaves as $\lambda^{-K}$, where $K$ will be estimated below. 

We assume that $D$ and $D^{-1}$ are analytical at $\lambda=\infty$ (i.e., $\det T_0 \neq 0$).
Considering coefficients by $\lambda^{-j}$ ($j=0,1,2,\ldots$) in the formula \rf{BIL},  we obtain the following invariants:
\be \bal  \label{fk}
 f_0 := u_0 \cdot v_0 \ , \ex
 f_1 := u_0 \cdot v_1 + u_1 \cdot v_0 \ , \ex
 f_2 := u_0 \cdot v_2 + u_1 \cdot v_1 + u_2 \cdot v_0 \ , \ex
 \dotfill \ex
 f_k := u_0 \cdot v_k + u_1 \cdot v_{k-1} + \ldots + u_k \cdot v_0 \ ,
\ea \ee
where $k < K$. In order to formulate a more precise statement, we define
\be 
    k_{max2} = \min \{ k_{max1}, k_{mn} \} \ ,
\ee 
where $k_{mn} = M + N -1 + \min \{ 0, 2 - N', 2 -M', 4 - M' -N' \}$ and $k_{max1}$ is given by \rf{kmax1}.

\begin{prop}  \label{bil-prop1}
Bilinear expressions $f_k$ ($k=0,\ldots,k_{max2}$), given by \rf{fk}, are preserved by the Darboux transformation (i.e., $\tilde f_k = f_k$) provided that $\det T_0 \neq 0$.
\end{prop}

\begin{rem}
If $M > N \geqslant 0$, $N' \leqslant N+2$, $M' \leqslant M+2$, 
then $k_{max2} = k_{max1}$.
\end{rem}

In some cases we can formulate stronger propositions. For $N' \leqslant 2$, $M' \leqslant 2$ (including the isospectral case) $k_{max2}$ in Proposition~\ref{bil-prop1} can be replaced  by
\be
 k'_{max2} = - 1 + \min \{ N, M, N+M \} \ .
\ee
If the normalization is canonical ($T_0 = I$) we can replace $k_{max2}$ by
\be
k''_{max2} = \min \{ k''_{max1}, k''_{mn} \} \ ,
\ee
where  $k''_{mn} = M+N + \min \{ 1, 2-N', 2-M', 3-M'-N' \}$.
\ods
Analogical considerations can be done for $\Tr U^2$ and $\Tr V^2$. To obtain the final results (see below) it is enough to 
substitue $M \rightarrow N$, $M' \rightarrow N'$ in the first case, and $N \rightarrow M$, $N' \rightarrow M'$  in the second case. 
\ods

\begin{prop}  \label{bil-prop2}
Suppose that $0 \leqslant k \leqslant k_{max3}$, where
\[
k_{max3} = \min \{ N-1, N+1-N', 2N-1, 2N+1-N', 2N+3-2N'  \}
\]
and $g_0, g_1,\ldots, g_k$ are given functions of $x$. Then the bilinear constraints 
\be \bal
 g_0 := u_0 \cdot u_0 \ , \ex
 g_1 := u_0 \cdot u_1 + u_1 \cdot u_0 \ , \ex
 g_2 := u_0 \cdot u_2 + u_1 \cdot u_1 + u_2 \cdot u_0 \ , \ex
 \dotfill \ex
 g_k := u_0 \cdot u_k + u_1 \cdot u_{k-1} + \ldots + u_k \cdot u_0 \ ,
\ea \ee
are preserved by the Darboux transformation such that $\det T_0 \neq 0$.
\end{prop}
\ods

\begin{prop}  \label{bil-prop3}
Suppose that $0 \leqslant k \leqslant k_{max4}$, where
\[
k_{max4} = \min \{ M-1, M+1-M', 2M-1, 2M+1-M', 2M+3-2M'  \}
\]
and $h_0, h_1,\ldots, h_k$ are given functions of $x$. Then the bilinear constraints 
\be \bal  \label{hk}
 h_0 := v_0 \cdot v_0 \ , \ex
 h_1 := v_0 \cdot v_1 + v_1 \cdot v_0 \ , \ex
 h_2 := v_0 \cdot v_2 + v_1 \cdot v_1 + v_2 \cdot v_0 \ , \ex
 \dotfill \ex
 h_k := v_0 \cdot v_k + v_1 \cdot v_{k-1} + \ldots + v_k \cdot v_0 \ ,
\ea \ee
are preserved by the Darboux transformation such that $\det T_0 \neq 0$.
\end{prop}

\subsection{Invariants for general Lax pairs}

Let us consider matrices $U$ and $V$ in the neighbourhood of $\lambda=\lambda_0$,
where $U, V$ have poles of $N$-th and $M$-th order, respectively, i.e., 
\be \label{Lax-Taylor}
U = \sum_{k=0}^\infty u_k (\lambda-\lambda_0)^{k - N}  \ ,
\quad
V = \sum_{k=0}^\infty v_k (\lambda-\lambda_0)^{k - M} \ , 
\ee
where $u_k = u_k (x)$, $v_k = v_k (x)$ ($k = 0, 1, 2, \ldots$). We are going to show that the general case reduces to the polynomial case discussed above. Indeed, it is sufficient to 
use the following parameter $z$ 
in the neighbourhood of $\lambda_0$:   
\be
    z = (\lambda - \lambda_0)^{-1} \ ,
\ee
and then the Lax pair 
\rf{Lax-Taylor} becomes identical with \rf{UV-poly}. Note that \ $z \rightarrow \infty$ \ for \ $\lambda \rightarrow \lambda_0$. 
We assume that the Darboux matrix $D$ is analytic at $\lambda_0$:
\[
    D = T_0 + T_1 (\lambda-\lambda_0) + T_2 (\lambda -\lambda_0)^2 + \ldots = 
 T_0 + z^{-1} T_1 + z^{-2} T_2 + \ldots 
\]
where matrices $T_k$ depend on $x$. 
In the nonisospectral case we transform the equations \rf{Lnu} to 
the form \rf{nonisy}:
\be
  z,_\nu = - z^2 L_\nu (x, \lambda_0 + z^{-1})  \ , 
\ee 
where $L_\nu$ have to be expanded in the Laurent (or Taylor) series at $z = \infty$. 

In order to obtain linear invariants we consider the linear combination of matrices $U, V$, given by:
\be
  U + (\lambda - \lambda_0)^{M-N} h V 
\ee
where
\be
h = h (x,y;\lambda) \equiv \sum_{k=0}^\infty (\lambda - \lambda_0)^k 
  h_k (x,y) = \sum_{k=0}^{\infty} h_k z^{-k} 
\ee
is a given scalar function, holomorphic at $\lambda = \lambda_0$. 
Finally, we arrive at an exact analogue of Proposition~\ref{lin-prop}. 

Bilinear invariants can be treated in the same way. 
We obtain exact analogues of Propositions~\ref{bil-prop1}, \ref{bil-prop2} and \ref{bil-prop3}. 

\begin{cor}
The polynomial case can be treated as a special subcase, defined by  
$\lambda_0 = \infty$. It is enough to change variables in formulas \rf{Lax-Taylor}:  
$\lambda \rightarrow \lambda^{-1}$ (and $\lambda_0 \rightarrow \lambda_0^{-1}$).  Then, making the limit $\lambda_0 \rightarrow 0$, we get  \rf{UV-poly}. 
\end{cor}

\subsection{Application to the KdV equation}

We will show advantages of Darboux invariants considering the case of the KdV equation. Our approach consists in characterizing the Lax pair in terms of some algebraic constraints (see \cite{Ci-ChSF,Ci-dbt}) and then showing that these constraints are preserved by the Darboux-B\"acklund transformation. 

\begin{prop}  \label{KdV-cons}
The Lax pair \rf{Lax-KdV} can be uniquely characterized by the following set of alebraic constraints:
\begin{enumerate}
\item $U$ is linear in $\lambda$ ($U = u_0 \lambda + u_1$), \quad $\Tr\,U = 0$ \ , 
\item $V$ is quadratic in $\lambda$ ($V = v_0 \lambda^2 + v_1 \lambda + v_2$), \quad  
$\Tr\,V = 0$ \ , 
\item  $u_0 = \mm 0 & 0 \\ -1 & 0 \ema$, \quad $u_1$ is off-diagonal. \label{3}
\item  \label{4}
$
u_0 - \frac{1}{4} v_0 = 0 \ , \quad u_1 - \frac{1}{4} v_1 \in g \ , \ 
$
where $g$ is the 1-dimensional Lie algebra generated by $\mm 0 & 0 \\ 1 & 0 \ema$, 
\item   \label{5}
$ u_0 \cdot v_1 + u_1 \cdot v_0 = - 8$ \ , \quad 
$v_1 \cdot v_1 + 2 v_0 \cdot v_2 = 0$ ,   
\item \label{6}
$\overline{ U (\lambda) } = U (\bar\lambda)$, \quad $\overline{ V (\lambda) } = V (\bar \lambda)$  . 
\end{enumerate}
\end{prop}

\no Proof: The first four properties imply the following form of $U, V$:
\be \ba{l} 
 u_0 = \mm 0 & 0 \\ -1 & 0 \ema \ , \quad u_1 = \mm 0 & p \\ u & 0 \ema \ , 
\\[3ex] 
v_0 = \mm 0 & 0 \\ -4 & 0 \ema \ , \quad  v_1 = \mm 0 & 4p \\ q & 0 \ema \ ,  
\quad v_2 = \mm -a & b \\ c & a \ema \ , 
\ea \ee
where $u, p, q, a, b, c$ are some complex fields.  Bilinear constraints \ref{5} yield
\be
 - 8p = -8 \ , \quad  8 p q - 8 b = 0 \ ,  
\ee
i.e., $p=1$, $q = b$. Now compatibility conditions yield the KdV equation \rf{KdV} and  expressions \rf{abc} for $a, b, c$. The last property implies $u \in \R$. \hfill $\Box$
\ods

The first two constraints are preserved by any Darboux transformation 
constructed in the standard way, e.g., using Corollary~\ref{cor-Dar} (and 
the tracelessness is preserved by virue of Remark~\ref{rem-detconst}, provided that $\det \N = \const$). The constraints~\ref{6} impose restrictions on $\lambda_k$ and $p_k$, see Section~\ref{sec-reality}.
In order to preserve the third constraint we have to use freedom in the choice of the normalization matrix $T_0 \equiv \N$. From the first equation of \rf{tilde-u01}  we get (taking into  account the form of $u_0$ given by the third constraint): 
\be  \label{NKdV} 
  \N =  f \mm 1 & 0 \\ \alpha  & 1 \ema  \ , 
\ee
where $f, \alpha$ are some functions.   In the sequel we put $f=1$ (thus $\det\N=1$). 
Then, denoting \ $T_1 = \mm c_1 & c_2 \\ c_3 & c_4 \ema$, we rewrite the second equation of \rf{tilde-u01} as
\be
    \mm 0 & 1 \\ \tilde u & 0 \ema = \mm - \alpha & 1 \\ u - \alpha^2 & \alpha \ema + 
\mm - c_2  & 0 \\ c_1 - \alpha c_2 - c_4 & c_2 \ema + \mm 0 & 0 \\ \alpha,_1 & 0 \ema  \ .
\ee
Hence:
\be  \label{alfac}
  \alpha = - c_2 \ , \quad \tilde u = u - \alpha^2 - \alpha c_2 + \alpha,_1 + c_1 - c_4 \ .  
\ee

\ods

The constraint~\ref{4} is preserved by virtue of Proposition~\ref{prop-g}. 
Other propositions from Section~\ref{invariants} are too weak for our present purposes. Indeed, in the KdV case we have 
$k_{max2} = 0$ and $k_{max4} = 1$. Therefore the preservation of the constraints \ref{5} does not follow from Propositions~\ref{bil-prop1} and \ref{bil-prop3}.

Fortunatelly, the special form of matrices $u_0, v_0$ and $T_0$ (given by \rf{NKdV} with $f=1$) for KdV equation allows us to reconsider the behaviour of leading terms  \rf{lead-2}.  
We easily see that any matrix product containing only matrices from the set $\{ T_0, T_0^{-1}, T_0,_1, T_0,_2, u_0, v_0 \}$ (and among them at least one matrix from the set $\{ T_0,_1, T_0,_2, u_0, v_0 \}$) is proportional to $u_0$, and, as a consequence, it has vanishing trace. Hence, in the case of the KdV equation the Darboux transformation preserves constraints \rf{fk} for $k=0,1$ and constraints \rf{hk} for $k=0,1,2$. In particular, the constraints~\ref{5} of Proposition~\ref{KdV-cons} are preserved. 

\begin{cor}
The Darboux transformation (defined as in Corollary~\ref{cor-Dar}) preserves all constraints defining the KdV Lax pair (see Proposition~\ref{KdV-cons}) 
provided that we impose reality restrictions on $\lambda_k$, $p_k$ (see Section~\ref{sec-reality}) and fix the normalization matrix according to the formula \rf{NKdV} where $f =1$ and $\alpha$ is expressed by the matrix $T_1$, namely $\alpha = - c_2$. 
\end{cor}

In the case of the elementary Darboux matrix $\det T_0 = 0$ and considerations presented in this section are not applicable. It would be interesting to 
extend the theory of Darboux invariants on the case $\det T_0 = 0$.

\section{Concluding remarks}

In this paper we gave a unified view on the Darboux-B\"acklund transformations for 
$1+1$-dimensional integrable systems of nonlinear partial differential equations.  
In particular, we discussed in detail relationships between various approaches to the construction of the Darboux matrix.  

Darboux-B\"acklund transformations have been extended in many directions. 
First of all, they are applicable to $2+1$-dimensional integrable systems \cite{BKP,DL,GHZ,MS,Ni}, 
including self-dual Yang-Mills equations \cite{GHZ,NGO,Us}. Then, we have $0+1$-dimensional systems, 
e.g., ordinary differential equations of nonlinear quantum mechanics \cite{CCU,DL,LC}. 
Darboux transformations were also constructed in the supersymmetric case \cite{LM-susy,Mi-susy} and in the non-commutative case \cite{SH}.

Matrix representations of spectral problems and Darboux transformations are not always convenient. Impressive examples  
are associated with Clifford algebras. 
It is enough to compare the paper \cite{Ci-izot}, where mainly the matrix approach was used, with subsequent papers \cite{BC-pla,Ci-jpal}, which are much shorter, more general and more elegant. All these papers consider binary Darboux transformation. An extension on multipole case is not so obvious, compare \cite{CB-jpal}, where some progress in this direction is described. There exist   other generalizations of the Darboux transformation on spectral problems with values in abstract associative algebras \cite{BdMM,DL}.  

The discrete case is (to some extent) very similar to the continuous case. Many aspects (e.g., those concerning the rational dependence on $\lambda$ and the loop group structure) are just repetitions from the continuous case, compare \cite{DL,LRB,Mat}. It is tempting to apply the ideas of time scales \cite{BoPet}, all the more so that in the ``classical'' case of the pseudospherical surfaces we succeded to construct the Darboux-B\"acklund transformation on arbitrary time scale \cite{Ci-time}, thus treating the discrete and continuous case in a uniform way. However, some points seem to be more difficult in the discrete case, e.g., Darboux invariants are not formulated yet. 
Actually, it is not so easy even to find an appropriate discretization of a given integrable system, 
especially if the associated linear problem is non-isospectral.

{\it Acknowledgements}: I am grateful to Maciej Nieszporski for many fruiful discussions.

\end{document}